%% file: main.tex
\documentclass[%
 reprint,
nofootinbib,
 amsmath,amssymb,
 aps,
 longbibliography,
]{revtex4-2}
\usepackage{graphicx}
\usepackage{dcolumn}
\usepackage{bm}

\usepackage{amsmath}
\usepackage{mathtools}
\usepackage{physics}
\input{ishibashi} 
\input{symbols}

\newcommand{\TQc}[1]{\textcolor{black}{#1}}
\newcommand{\JHc}[1]{\textcolor{black}{#1}}
\newcommand{\JHcc}[1]{\textcolor{black}{#1}}

\usepackage{tikz}
\usetikzlibrary{calc
, decorations.pathreplacing
, decorations.markings
, arrows}

\pgfmathsetmacro\MathAxis{height("$\vcenter{}$")}

\usepackage{hyperref}
\usepackage[capitalize]{cleveref}
\begin{document}

\preprint{APS/123-QED}

\title{Revisiting the symmetry-resolved entanglement for non-invertible symmetries in $1{+}1$d conformal field theories}

\author{Jared Heymann}
 \email{j.heymann@unimelb.edu.au}
\author{Thomas Quella}%
 \email{Thomas.Quella@unimelb.edu.au}
\affiliation{%
The University of Melbourne\\School of Mathematics and Statistics\\Parkville 3010 VIC, Australia
}%

\date{\today}

\begin{abstract}
  Recently, a framework for computing the symmetry-resolved entanglement entropy for non-invertible symmetries in $1{+}1$d conformal field theories has been proposed by Saura-Bastida, Das, Sierra and Molina-Vilaplana [\href{https://doi.org/10.1103/PhysRevD.109.105026}{Phys.\,Rev.\,D\textbf{109},\,105026}]. 
  We revisit their theoretical setup, paying particular attention to possible contributions from the conformal boundary conditions imposed at the entangling surface -- a potential subtlety that was not addressed in the original proposal. We find that the presence of boundaries modifies the construction of projectors onto irreducible sectors, compared to what can be expected from a pure bulk approach. This is a direct consequence of the fusion algebra of non-invertible symmetries being different in the presence or absence of boundaries on which defects can end.

  We apply our formalism to the case of the Fibonacci category symmetry in the three-state Potts and tricritical Ising model and the Rep($S_3$) fusion category symmetry in the $SU(2)_4$ Wess-Zumino-Witten conformal field theory. 
  We numerically corroborate our findings by simulating critical anyonic chains with these symmetries.
  Our predictions for the symmetry-resolved entanglement for non-invertible symmetries seem to disagree with the recent work by Saura-Bastida et al.
\end{abstract}

\maketitle


\section{Introduction}
Entanglement plays a key role in both understanding and simulating quantum many-body systems \cite{laflorencieQuantumEntanglementCondensed2016}. It is used to identify phenomena such as quantum phase transitions and critical points \cite{vidalEntanglementQuantumCritical2003, osterlohScalingEntanglementClose2002,Korepin:2004PhRvL..92i6402K, calabreseEntanglementEntropyConformal2009,Tu:2017PhRvL.119z1603T}, and to characterise phases that lack local order parameters, such as topological phases \cite{levinDetectingTopologicalOrder2006, kitaevTopologicalEntanglementEntropy2006}. In the context of numerical approaches, the success of tensor network algorithms \cite{whiteDensityMatrixFormulation1992, sierraEquivalenceVariationalMatrix1998} hinges on the entanglement structure displayed by ground states \cite{Hastings:2007PhRvB..76c5114H,eisertAreaLawsEntanglement2010}: understanding and controlling entanglement allows more efficient simulations of quantum many-body systems \cite{schollwoeckDensitymatrixRenormalizationGroup2011, banulsTensorNetworkAlgorithms2023}.

Recently, there has been a surge of interest in understanding the interplay between entanglement and another pillar of modern physics: symmetries.
Of particular interest has been the so-called symmetry resolved entanglement entropy (SREE), which quantifies the entanglement entropy in the different representations of a symmetry.
The idea of symmetry resolution dates back to Ref.~\cite{laflorencieSpinresolvedEntanglementSpectroscopy2014}, where the entanglement spectra of critical spin chains was resolved with respect to the spin quantum number. 
More recently, the SREE has been studied extensively in $ 1{+}1$d conformal field theories (CFTs) due to the powerful analytical methods available \cite{goldsteinSymmetryresolvedEntanglementManybody2018, xavierEquipartitionEntanglementEntropy2018, bonsignoriSymmetryResolvedEntanglement2019, capizziSymmetryResolvedEntanglement2020, bonsignoriBoundaryEffectsSymmetry2021, calabreseSymmetryresolvedEntanglementEntropy2021, digiulioBoundaryConformalField2023, kusukiSymmetryresolvedEntanglementEntropy2023, northeEntanglementResolutionRespect2023, Banerjee:2024ldl}
(for a more comprehensive review see Ref.~\cite{Castro-Alvaredo:2024azg}
The SREE has also found applications in studying strongly correlated models \cite{lukinProbingEntanglementManybodylocalized2019, shenDisentanglingPhysicsAttractive2024} where it can provide non-trivial information concerning the configuration of the ground state or serve as an order parameter \cite{araEntanglementEdgeModes2024}.
From the numerous studies of the SREE in $1{+}1$d CFTs, a key result has emerged: the equipartition of the SREE.
 At leading order in the ultraviolet cutoff, the SREE is equally distributed among the different representations of the theory.
 Violations of the equipartition are observed at subleading order and are generally sensitive to the dimension of the representation in question \cite{goldsteinSymmetryresolvedEntanglementManybody2018,calabreseSymmetryresolvedEntanglementEntropy2021, kusukiSymmetryresolvedEntanglementEntropy2023}.
 
 A powerful approach to symmetry resolution in $ 1{+}1$d CFTs adopts the framework of boundary CFT (BCFT)
 \cite{cardyBoundaryConditionsFusion1989, digiulioBoundaryConformalField2023, northeEntanglementResolutionRespect2023, kusukiSymmetryresolvedEntanglementEntropy2023}.
 The BCFT approach to the SREE arises because the spectra of the entanglement Hamiltonian in $ 1{+}1$d CFTs is described by a BCFT \cite{cardyEntanglementHamiltoniansTwodimensional2016, lauchliOperatorContentRealspace2013, ohmoriPhysicsEntanglingSurface2015}.
 A BCFT is present because computing the entanglement entropy requires a bipartition of the Hilbert space, which comes with a choice of boundary conditions at the entangling surface \cite{ohmoriPhysicsEntanglingSurface2015}.
 The choice of boundary conditions dictates the energy spectrum of the entanglement Hamiltonian \cite{cardyEntanglementHamiltoniansTwodimensional2016} and which symmetries of the bulk are present in the BCFT \cite{kusukiSymmetryresolvedEntanglementEntropy2023, choiRemarksBoundariesAnomalies2023}.
 By using the BCFT approach to the SREE, when the symmetry in the BCFT is a finite group $ G$, Ref.~\cite{kusukiSymmetryresolvedEntanglementEntropy2023} found\footnote{For a derivation of this result within the framework of algebraic quantum field theory, see Refs.~\cite{Casini:2019kex,Magan:2021myk}}
 \begin{align}
   \lim_{q\to 1} \bigl[S_n(q,r)- S_n(q)\bigr] = \log \frac{d_r^2}{\abs{G} }\;, \label{eq:equipartition-group}
 \end{align}
 where $ d_r$ is the dimension of the representation $ r$ of $ G$ and $ \abs{G} $ is the cardinality $ G$.
 Here, $ S_n(q,r)$ is the $ n^{\text{th}}$ R\'enyi entropy within the irreducible representation $ r$, while $ S_n(q)$ is the $ n^{\text{th}}$ R\'enyi entropy.
 The $ q \rightarrow 1$ limit corresponds to taking the ultraviolet cutoff (lattice spacing) to $ 0$ (or the width of the annulus in the BCFT to infinity).
 When the symmetry of interest is a finite group $G$, \cref{eq:equipartition-group} indicates that the equipartition of the SREE is broken by a term that depends on the dimension of the $ r$ representation of $G$ and the cardinality of $G$.
 
 In recent years, after the seminal work of Ref.~\cite{gaiottoGeneralizedGlobalSymmetries2015} the notion of a symmetry in quantum field theory and quantum lattice models has been generalised systematically and now extends beyond the concept of groups. 
 As part of this endeavour, there have also been substantial efforts to abstract the concept of the symmetry (and its representations) from that of the underlying physical system in terms of what has become to be known as a ``SymTFT'' or sandwich construction. 
 For recent reviews on these very dynamical areas of research and references to the original literature see Refs.~\cite{Bhardwaj:2018JHEP...03..189B, mcgreevyGeneralizedSymmetriesCondensed2023,Schafer-Nameki:2023jdn,Bhardwaj:2023arXiv230707547B,shaoWhatDoneCannot2023,Carqueville:2023arXiv231102449C}.
 From the modern perspective that has emerged, symmetries should be identified with topological defects of various dimensions and composition of symmetries be realised as defect fusion. 
 In this framework it is very natural to allow for symmetries that are not invertible.
 Predating the more recent efforts, the consideration of topological defects and their role in terms of describing symmetries and dualities has a long history in the study of $1{+}1$d rational CFTs (RCFTs), see e.g. Refs~\cite{petkovaGeneralizedTwistedPartition2001, frohlichKramersWannierDualityConformal2004,frohlichDualityDefectsRational2007}.
 
 Recently, in Ref.~\cite{saura-bastidaCategoricalsymmetryResolvedEntanglement2024} the SREE was analysed for non-invertible symmetries, dubbed the Cat-SREE, generated by Verlinde lines in $ 1{+}1$d RCFTs.
 The tricritical Ising model was studied and the entanglement entropy was resolved with respect to the Fibonacci category.
 The Fibonacci category has two simple objects $C= \{1, W\}$ with non-trivial fusion relation $ W \times W = 1 + W$ and two irreducible representations, also labelled by $C$.
 Analogous to the group-like case, Eq.~(33) of~Ref.~\cite{saura-bastidaCategoricalsymmetryResolvedEntanglement2024} implies\footnote{A similar result has been derived using the framework of algebraic quantum field theory \cite{Benedetti:2024dku}.} 
 \begin{align}
    \lim_{q \to 1}\left[ S_n(q,r) - S_n(q) \right] = \log \frac{d_r^2}{\abs{C}^2 } \label{eq:equipartition-wrong}
 \end{align}
 where $ d_r$ is the (quantum) dimension of the representation $r \in C$ and $ \abs{C}^2 = \sum_{c \in C}^{} d_c^2 $.\footnote{We note that our convention for the total quantum dimension $\abs{C}$ differs from the one used in \cite{saura-bastidaCategoricalsymmetryResolvedEntanglement2024}.}
 However, we find that the above breaking of equipartition \labelcref{eq:equipartition-wrong} is incorrect.
 \JHc{Instead, for symmetries generated by self-dual, multiplicity free Verlinde lines and Cardy boundary conditions $a$ and $b$ imposed at the entangling surfaces, we find that the equipartition of the SREE when projecting onto the $r$ boundary conformal tower is}
 \begin{align}
    \lim_{q \to 1}\left[ S_n(q,r) - S_n(q) \right]=\log \frac{d_r}{d_a d_b}\;. \label{eq:equipartition-correct2}
 \end{align}
\JHc{
If we choose both boundaries to be identical, $a = b =a_C$, with $a_C$ a boundary symmetric under $C$ such that $  a_C \times a_C = 
 \sum_{c \in C} c$,  then \cref{eq:equipartition-correct2} implies that the equipartition of the SREE is}
 \begin{align}
    \lim_{q \to 1}\left[ S_n(q,r) - S_n(q) \right]=\log \frac{d_r}{\sqrt{C} }\;, \label{eq:equipartition-correct}
 \end{align}
 where $\sqrt{C} \coloneq \sum_{c \in C} d_c$ and $r$ labels an irreducible representation of $C$.
 When the Verlinde lines form a finite (abelian) group, \cref{eq:equipartition-correct} is in agreement with \cref{eq:equipartition-group} because  the quantum dimensions of invertible symmetries is equal to one.
 The deviation from Ref.~\cite{saura-bastidaCategoricalsymmetryResolvedEntanglement2024} can be succinctly summarised as follows: \emph{the fusion algebra of non-invertible symmetries is sensitive to the presence of boundary conditions and is different than the fusion algebra of the bulk.}
 In this work, we find that if one wishes to understand the relation between entanglement measures and non-invertible symmetries, extra care due to the presence of boundaries is required -- an issue that was not addressed in Ref.~\cite{saura-bastidaCategoricalsymmetryResolvedEntanglement2024}.

\JHcc{
\cref{eq:equipartition-correct} requires that our multiplicity free fusion category contains a self-dual object that generates all simple objects of the fusion category $C$.
These are rather stringent conditions.
Therefore, we emphasize that \cref{eq:equipartition-correct} is not a general formula and we can only show its validity for fusion categories that satisfy the conditions mentioned above. To mention a specific example, \cref{eq:equipartition-correct} is valid for $SU(2)_k$ fusion categories restricted to their integer spin Verlinde lines; the restriction to integer spin Verlinde lines always results in a symmetric boundary state \cite{choiRemarksBoundariesAnomalies2023}.
Moreover, it is straightforward to show that the symmetric boundary state(s) also generates all integer spin Verlinde lines.\footnote{For even (odd) $k$, the boundary state(s) with spin label(s) $ a_C = \frac{k}{4}$ $(\frac{k\pm1}{4})$ satisfies $a_C \times a_C = \sum_{c\in C} c$ where $C$ is the $SU(2)_k$ fusion category restricted to integer spin labels.}
Small values of $k$ result in familiar fusion categories: the Ising category ($k=2$), the Fibonacci category ($k=3$) and the $\mathrm{Rep}(S_3)$ category ($k=4$) \cite{choiRemarksBoundariesAnomalies2023}.
}

  Before proceeding to more technical considerations, we note that topological defects in the presence of boundaries find numerous other applications. In the CFT setting, they have been used to analyse boundary renormalisation group flows \cite{grahamDefectLinesBoundary2004, konechnyOpenTopologicalDefects2020} and study open string field theory \cite{kojitaTopologicalDefectsOpen2018}. The algebraic structure formed by considering topological defects terminating on two domain walls (or boundaries) is known as the ``ladder algebra'' in the mathematics literature \cite{Barter:2018hjs, 2019Rowett, 2023Henriques};
 this has its roots in analysing gapped boundaries and domain walls for topological phases \cite{Kitaev:2011dxc}.
 However, the ladder algebra can also be viewed analogously to the tube algebra in the bulk \cite{linAsymptoticDensityStates2023}, providing the appropriate mathematical setting to analyse generalised symmetries in the presence of boundaries.
 This has motivated a recent SymTFT perspective of the ladder algebra to study the representation theory of solitons \cite{Copetti:2024dcz, Copetti:2024onh,cordovaRepresentationTheorySolitons2024}. We would also like to mention Ref.~\cite{Vanhove:2021nav}, where the ladder algebra was discussed in the context of the three-state Potts model.
 
 The rest of this paper is organised as follows.
 In \cref{sec:SREE-BCFT-review} we briefly review the BCFT approach to computing the SREE.
 In \cref{sec:Counter-Example} we provide an explicit counter example to the results in Ref.~\cite{saura-bastidaCategoricalsymmetryResolvedEntanglement2024} by analysing the SREE with respect to the Fibonacci category in the three-state Potts model.
 In \cref{sec:TDL} we briefly review topological defect networks and the action of defects on states in the open string Hilbert space.
 This allows us to construct projections to individual symmetry sectors on the open string Hilbert space and analyse the asymptotic limit of the SREE in \cref{sec:Constructing-Projectors}.
 A crucial ingredient is a detailed understanding of the fusion algebra of defects in the presence of boundaries. We then proceed to apply our framework to specific examples in \cref{sec:MFC-SREE-examples} and numerically corroborate our results by simulating anyonic chains.
 
 \section{Brief review of the BCFT approach to the SREE} \label{sec:SREE-BCFT-review}
\begin{figure}[]
   \centering
   \includegraphics[width=\columnwidth]{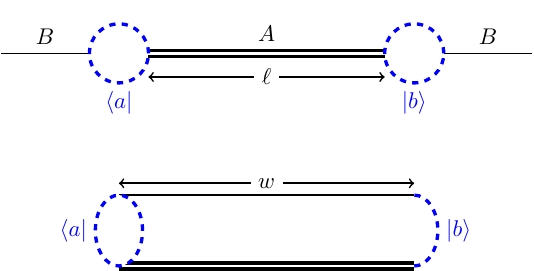}
   \caption{Two slits of radius $ \varepsilon$ are inserted at the entangling surface and boundary conditions $ a$ and $ b$ are imposed.
  By a conformal transformation, the region $ A $ is mapped to an annulus of width $ w$ with boundary states $ \ket{a}$ and $ \ket{b}$.}
   \label{fig:map-to-annulus}
 \end{figure}

\JHc{This brief review closely follows the formalism established in Ref.\ \cite{kusukiSymmetryresolvedEntanglementEntropy2023}.} 
 Defining the entanglement between regions $A$ and $B$ for a state described by the density matrix $\rho$ requires a bipartition of the Hilbert space 
 \begin{align}
     \mathcal{H} = \mathcal{H}_A \otimes \mathcal{H}_B\;. \label{eq:bipartition-1}
 \end{align}
 The reduced density matrix $\rho_A = \tr_B \rho$ corresponding to region $A$ is obtained by tracing out the degrees of freedom in $B$, and the entanglement entropy can be computed.
 However, as explained in Ref.~\cite{ohmoriPhysicsEntanglingSurface2015}, a well defined bipartition of the total Hilbert space \labelcref{eq:bipartition-1} requires a choice of boundary conditions at the entangling surface (the regions separating $ A$ and $ B$). 
 The need for boundary conditions at the entangling surface is most explicit in theories whose Hilbert space does not admit a simple tensor product decomposition. \TQc{These include continuum field theories but also lattice models} such as anyonic chains \cite{feiguinInteractingAnyonsTopological2007} which in turn represent important examples of physical systems with non-invertible symmetries.
 For a conformal field theory, it is natural to choose boundary conditions that preserve the (extended) chiral algebra.
 Formally, this is achieved by a factorisation map $ \iota_{a,b}$:
 \begin{align}
   \iota_{a,b}: \mathcal{H} \rightarrow \mathcal{H}_{A,ab} \otimes \mathcal{H}_{B,ab}\;,
 \end{align}
 where $ \mathcal{H}_{A,ab}$ ($ \mathcal{H}_{B,ab}$) is the Hilbert space in region $ A$ ($B$) with boundary conditions $ a$ and $ b$ at the entangling surface.
 In the path integral, the factorisation map is implemented by inserting two disks of radius $ \varepsilon \ll 1$ that serve as an ultraviolet (UV) cutoff.
 The region $ A$ of length $ \ell$ is then mapped to an annulus of width $ w$ by a conformal transformation \cite{ohmoriPhysicsEntanglingSurface2015,cardyEntanglementHamiltoniansTwodimensional2016}, see \cref{fig:map-to-annulus}.
 The dependence of $ w$ on $ \ell$ depends on the initial geometry of the problem: for instance, whether the Hilbert space is defined on an infinite line or finite periodic chain etc..
 For a comprehensive list of different geometries see Ref.~\cite{cardyEntanglementHamiltoniansTwodimensional2016}.
 For our purposes, we are interested in the BCFT defined on an annulus of width $ w$ and it is not important what the initial geometry of the problem was.
 
 The reduced density matrix $ \rho_{A,ab} = \tr_{B,ab} \rho$ is obtained by tracing out the degrees of freedom in $ \mathcal{H}_{B,ab}$, where $ \rho$ is the density matrix \TQc{corresponding to the} ground state. \TQc{It was argued in \cite{cardyEntanglementHamiltoniansTwodimensional2016} that the reduced density matrix can be written as the annulus partition function of an appropriate boundary CFT. More specifically,}
 \begin{align}
   \rho_{A,ab} &= \frac{q^{L_0 - c/24}}{Z_{ab}(q)}\;, \label{eq:rdm}
 \end{align}
 where $ q = e^{-2 \pi^2/w} $ is the modular parameter, $ c$ is the central charge,
 \begin{align}
   Z_{ab}(q) = \tr_{ab}q^{L_0 - c/24} \;,
 \end{align}
 is the open string partition function with boundaries $ a$ and $ b$ and $ \tr_{ab}$ denotes the trace over the open string Hilbert space with boundaries $ a$ and $ b$.
 The normalisation factor of $ 1/Z_{ab}(q)$ in \cref{eq:rdm} ensures $ \tr_{ab} \rho_{A,ab} = 1$.
 
  By a modular transformation, the open string partition function is written in the closed string channel as
 \begin{align}
   Z_{ab}(q) =  \bra{a} \tilde{q}^{L_0 - c/24} \ket{b}\;,
 \end{align}
 where $ \tilde{q} = e^{-2 w}$ is the dual modular parameter and $ \ket{a}$ and $ \ket{b}$ are conformal boundary states imposed at the entangling surface by the factorisation map $ \iota_{ab}$.
 The R\'enyi entropies are 
 \begin{align}
   S_n(q)\coloneq \frac{1}{1-n} \log \tr_{ab} \rho_A^n = \frac{1}{1-n}\log \frac{Z(q^n)}{Z^n(q)}\;,
 \end{align}
 with the entanglement entropy being given by $ \lim_{n \rightarrow 1} S_n(q)$.

  \TQc{The framework just described was used and extended systematically in Ref.~\cite{kusukiSymmetryresolvedEntanglementEntropy2023} to allow for the consideration of symmetry resolution of these entropies. While the previous paper was concerned with continuous or finite group symmetries we will suppose now that}
  we have some symmetry category $ C$ acting on the open string Hilbert space. \TQc{And just as in Ref.~\cite{kusukiSymmetryresolvedEntanglementEntropy2023} we assume that this action corresponds to the one inherited from the reduction of the action of $ C$ on the total Hilbert space \eqref{eq:bipartition-1} to that on the reduced space $\mathcal{H}_{A,ab}$.}
 The SREE is \TQc{then} computed by introducing 
 \begin{align}
   \mathcal{Z}_{ab}(q^n, r) \coloneq \tr_{ab} \Pi_r \rho_A^n\;, \label{eq:probabilities}
 \end{align}
 where $ \Pi_r$ projects onto the representation $ r$ of $ C$. 
 The symmetry resolved R\'enyi entropy (SRRE) is given analogously to the R\'enyi entropies
 \begin{align}
   S_n(q,r) \coloneq \frac{1}{1-n} \log \frac{\mathcal{Z}_{ab}(q^n,r)}{\mathcal{Z}_{ab}^n(q,r)}\;, \label{eq:SRRE}
 \end{align}
 with the SREE being obtained in the limit $n \to 1$.
 It is useful to define the charged moments of the reduced density matrix 
 \begin{align}
   z_{ab}(q^n,c) \coloneq \tr_{ab} \mathcal{L}_c q^{n(L_0 - c/24)}\;,  \label{eq:charged-moments}
 \end{align}
 where $ \mathcal{L}_c$ is the operator corresponding to the object $ c \in C$ acting on states in the open string Hilbert space.
 In the closed string channel the charged moments read 
 \begin{align}
   z_{ab}(q^n ,c) = \prescript{}{c}{\bra{a}} \tilde{q}^{\frac{1}{n}(L_0 - c/24)} \ket{b}_c\;, \label{eq:charged-moments-close}
 \end{align}
 where $ \ket{a}_c$ and $ \ket{b}_c$ are boundary states in the Hilbert space of the $ c$-twisted sector.
 The charged moments \labelcref{eq:charged-moments} correspond to computing the open string partition function with an insertion of the defect $ \mathcal{L}_c$ terminating on the boundaries $ a$ and $ b$.
 We require that $ \mathcal{L}_c$ can topologically end on both $ \ket{a}$ and $ \ket{b}$.
 This requires that the fusion $ c \times a$ contains $ a$ and, likewise, the fusion $ c \times b$ contains $ b$.
 Using the terminology of Ref.~\cite{choiRemarksBoundariesAnomalies2023}, the boundaries $ a$ and $ b$ must be weakly symmetric under $ \mathcal{L}_c$ for \cref{eq:charged-moments} to be non-zero.
 Moreover, we will assume that $ a$ and $ b$ are simple boundaries; they are not written as a superposition of elementary boundary states.
 
 The charged moments in the closed string channel make it quite simple to derive the asymptotic behaviour of the SREE.
 Generally, a projector onto the irreducible representation $ r$ of $ C$ is written as a linear combination of operators in $ C$,
 \begin{align}
   \Pi_r^{[a,b]} = \sum_{c\in C}^{} \Lambda_{r c}^{[a,b]} \mathcal{L}_c\;, \label{eq:projector-ansatz}
 \end{align}
 for constants $ \Lambda_{rs}^{[a,b]}$ that ensure the projectors are orthogonal.
 The superscripts $ [a,b]$ make it explicit that such projectors are defined in the presence of boundaries $ a$ and $ b$.
 If the boundaries are equal, $ a = b$, we will only include one boundary in the superscript, $[a]\equiv[a,a] $.
 For the symmetry $ C$ to be present in the open string Hilbert space, all defects $\mathcal{L}_c$, for $ c \in C$, must be able to topologically terminate on the boundaries $ a$ and $ b$: the boundaries $ a $ and $ b $ must be weakly symmetric under $ C$.
 As a consequence, $C$ is non-anomalous and be ``gauged'' in a generalised sense \cite{choiRemarksBoundariesAnomalies2023}.
 
 Using the general form of a projector \labelcref{eq:projector-ansatz}, in the closed string channel \cref{eq:probabilities} reads
 \begin{align}
 \mathcal{Z}_{ab}(q^n, r) &= \sum_{c \in C}^{} \Lambda_{rc}^{[a,b]} \prescript{}{c}{\bra{a}}\tilde{q}^{\frac{1}{n}(L_0 - c/24)}\ket{b}_c\;, \nonumber \\
     &\stackrel{q \rightarrow 1}{\sim} \Lambda_{r1}^{[a,b]} \tilde{q}^{\frac{-c}{24n}} \bra{a}\ket{0}\bra{0}\ket{b} \;,\label{eq:asymptotic-general}
 \end{align}
 where we have assumed that there is a unique vacuum $ \ket{0}$ in the bulk CFT.
 Therefore, in the asymptotic limit $ q \rightarrow 1$, the vacuum sector dominates in the closed string channel, leading to \cref{eq:asymptotic-general}.
 It will be useful to define the difference between the symmetry resolved R\'enyi entropies and the R\'enyi entropies
 \begin{align}
   \Delta S_n(q,r) \coloneq S_n(q,r) - S_n(q)\;. \label{eq:SREE-diff}
 \end{align}
 Plugging in the symmetry resolved R\'enyi entropies \labelcref{eq:SRRE} and utilising the asymptotic limit \labelcref{eq:asymptotic-general}, one finds that the equipartition of the SREE is broken by the term
 \begin{align}
   \lim_{q\to 1} \Delta S_n(q,r) = \log \Lambda_{r1}^{[a,b]}\;. \label{eq:SREE-equipartition}
 \end{align}
 For a finite group $ G$, the projectors onto the $r $ irrep are 
 \begin{align}
   P^{\mathrm{G} }_r &= \frac{\chi_r(1)}{\abs{G} }\sum_{g\in G}^{}  \overline{\chi_r(g)} U(g)\;,
 \end{align}
 where $ U(g)$ is the representation of the element $ g \in G$ and $ \chi_r(g)$ is the group character.
 From the above projector, one has $ \Lambda_{r1}^{[a,b]} = \chi_r(1)^2/\abs{G} $, recovering \cref{eq:equipartition-group}.
 
 \section{Symmetry resolution for the Fibonacci category}  \label{sec:Counter-Example}
 
 In this section we will revisit the Cat-SREE framework proposed in Ref.~\cite{saura-bastidaCategoricalsymmetryResolvedEntanglement2024} and show it is insufficient to produce the correct SREE with respect to the Fibonacci category in the three-state Potts model.
 Using the extended $ W_3$ algebra, the three-state Potts model has a diagonal modular invariant.
 There are six primary fields $ 1, \psi, \psi^\dagger, \varepsilon, \sigma$ and $\sigma^\dagger$ with scaling dimension $ 0,\; 2/3,\;2/3,\; 2/5,\; 1/15 $ and $ 1/15$, respectively.
 In the basis $ (1, \varepsilon, \psi, \sigma , \psi^\dagger, \sigma^\dagger)$ the modular $ S$ matrix reads \cite{difrancescoConformalFieldTheory1997}
 \begin{align}
   S &= \frac{1}{\sqrt{3} }\mqty(
   s & s & s \\
   s & \omega s& \omega^2 \\
   s & \omega^2 s & \omega s
   )\;,
 \end{align}
 with 
 \begin{align}
   s &= \frac{2}{\sqrt{5} }\mqty(
   \sin \frac{\pi}{5} & \sin \frac{2\pi}{5} \\
   \sin \frac{2 \pi}{5} & -\sin \frac{\pi}{5}
   )\;, \qquad \omega = e^{2\pi i/3} \;.
 \end{align}
 The fusion rule 
 \begin{align}
   \varepsilon \times \varepsilon = 1 + \varepsilon\;
 \end{align}
 indicates that the Verlinde lines $ \mathcal{L}_c$, for $ c\in C = \{1, \varepsilon \}$, generate a Fibonacci symmetry.
 For the case of diagonal RCFTs, the primary operators, Verlinde lines and boundary states are in one-to-one correspondence.
 Therefore, they can all share labels in some modular fusion category $ \mathcal{M} $.
 Thus, the Cardy state $ \ket{ \varepsilon}$ is weakly symmetric  with respect to $C$ \cite{choiRemarksBoundariesAnomalies2023}. The open string partition function with both boundaries $ a = b = \varepsilon$ is 
 \begin{align}
   Z_{\varepsilon \varepsilon} = \chi_1(q) + \chi_\varepsilon(q)\;.
 \end{align}
 
 Next, we require projectors onto the irreducible representations of $ C$.
 As $ C$ has two elements there are two irreducible representations.
 In the bulk CFT, the action of the Verlinde line $ \mathcal{L}_\varepsilon$ on each conformal family \cite{petkovaGeneralizedTwistedPartition2001} in the basis $ (1, \varepsilon, \psi, \sigma ,\psi^\dagger ,\sigma^\dagger)$ is
 \begin{align}
   \mathcal{L}_\varepsilon &= \qty( \frac{S_{\varepsilon j}}{S_{1j}}) =\mqty( \varphi,& 1- \varphi, & \varphi, & 1- \varphi,& \varphi, & 1- \varphi)\;,
 \end{align}
 where $\varphi = (1+\sqrt{5})/2$ is the golden ratio.
 The action of the topological defect $\mathcal{L}_c$ on the vacuum representation is given by the quantum dimension $d_c$ of $c$.
 This means that the conformal towers  $ 1,\, \psi$ and  $\psi^\dagger$ transform trivially under $ \mathcal{L}_\varepsilon$, while the conformal towers $ \varepsilon ,\, \sigma$ and $ \sigma^\dagger $ transform non-trivially under $ \mathcal{L}_\varepsilon$.
 
 For the bulk CFT with diagonal modular invariant, the Hilbert space decomposes as $ \mathcal{H}= \bigoplus_\mu \mathcal{V}_\mu \otimes \overline{\mathcal{V}}_\mu$.
 In the bulk CFT, the projectors $ P_r^{\mathrm{bulk} }: \mathcal{H} \rightarrow \mathcal{V}_r \otimes \overline{\mathcal{V}}_r$ onto the conformal tower $ r$ are \cite{linAsymptoticDensityStates2023}
 \begin{align}
   P_r^{\mathrm{bulk} } &= \sum_{b \in \mathcal{M} }^{} S_{1r} \overline{S}_{b r}\; \mathcal{L}_b\;. \label{eq:MFC-projector}
 \end{align}
 The primary operators $ \{1, \psi, \psi^\dagger \}$ transform in the trivial representation of $C$.
 Therefore, the projector onto the trivial irrep of $C$ is the sum of projectors onto the conformal towers $ \{1, \psi, \psi^\dagger \}$,
 \begin{align}
   \Pi_1^{\mathrm{bulk} } &= P_1^{\mathrm{bulk} } + P_\psi^{\mathrm{bulk} } + P_{\psi^\dagger}^{\mathrm{bulk} }\;.
 \end{align}
 Similarly, the primary operators $ \{\varepsilon, \sigma, \sigma^\dagger \}$ transform in the non-trivial irrep of $C$; the projector onto the non-trivial irrep of $C$ is the sum of projectors onto the conformal towers $ \{\varepsilon, \sigma, \sigma^\dagger \}$,
 \begin{align} 
   \Pi_\varepsilon^{\mathrm{bulk} } &= P_\varepsilon^{\mathrm{bulk} } + P_\sigma^{\mathrm{bulk} } + P_{\sigma^\dagger}^{\mathrm{bulk} }\;.
 \end{align}
 Using the bulk modular fusion category projectors \labelcref{eq:MFC-projector} we find
 \begin{align}
   \Pi_1^{\mathrm{bulk} } &= \frac{1}{1+\varphi^2}\qty( \mathcal{L}_1 + \varphi \mathcal{L}_\varepsilon) \\
   \Pi_\varepsilon^{\mathrm{bulk} } &= \frac{\varphi}{1+\varphi^2}\qty(\varphi \mathcal{L}_1 - \mathcal{L}_\varepsilon)\;. \label{eq:Fib-projectors}
 \end{align}
 As observed in Ref. \cite{saura-bastidaCategoricalsymmetryResolvedEntanglement2024}, these are the projectors one would obtain when using \cref{eq:MFC-projector} with the modular $ S$ matrix for the Fibonacci category,
 \begin{align}
   S_{\rm{Fib}} &= \frac{1}{\sqrt{1+\varphi^2} }\mqty(1 & \varphi \\\varphi & -1)\;.
 \end{align}
 
 We now need to compute the charged moment $ z_{\varepsilon\varepsilon}(q, \varepsilon)$, which requires computing the overlap of twisted boundary states. 
 The Hilbert space of the $ c$-twisted sector decomposes as \cite{kusukiSymmetryresolvedEntanglementEntropy2023}
 \begin{align}
   \mathcal{H}_{\mathcal{L}_c} &= \bigoplus_{\mu,\nu} N_{c \nu}^\mu \mathcal{V}_\mu \otimes \overline{\mathcal{V}}_\nu\;.
 \end{align}
 This implies that the diagonal sectors in $ \mathcal{H}_{\mathcal{L}_\varepsilon}$ are of the form $ \mathcal{V}_\rho \otimes \overline{\mathcal{V}}_\rho$ with $ \rho$ running over the set $ \{\varepsilon, \sigma ,\sigma^\dagger\}$.
 We denote the boundary state $ \ket{ \varepsilon}$ in the $ \varepsilon$-twisted sector by 
 \begin{align}
   \ket{\varepsilon}_\varepsilon &=a_1 \ishiket{\varepsilon}_\varepsilon + a_2 \ishiket{\sigma}_\varepsilon + a_3 \ishiket{\sigma^\dagger}_\varepsilon\;,
 \end{align}
   where  $\ishiket{h}_\varepsilon$ are the Ishibashi states in the $\varepsilon$-twisted sector and $ a_i$ are undetermined constants.
 Using a vector notation for the characters, this implies the charged moment $z_{\varepsilon \varepsilon}(q, \varepsilon) = \tr_{\varepsilon \varepsilon} \mathcal{L}_\varepsilon \rho_{A,\varepsilon, \varepsilon}
 $
 is  
 \begin{align}
   z_{\varepsilon \varepsilon}(q, \varepsilon) &= {\alpha}_1 \chi_\varepsilon(\tilde{q}) + \alpha_2 \chi_\sigma(\tilde{q}) + \alpha_3 \chi_{\sigma^\dagger}(\tilde{q}) \nonumber \\
   &=    \mqty(
   \mu (\alpha_1 + \alpha_2 + \alpha_3) \\
   \nu (\alpha_1+\alpha_2 +\alpha_3) \\
   \mu(\alpha_1 + \omega^2 \alpha_2 + \omega \alpha_3) \\
   \nu (\alpha_1 + \omega^2 \alpha_2 + \omega \alpha_3) \\
   \mu(\alpha_1 + \omega \alpha_2 + \omega^2 \alpha_3) \\
   \nu(\alpha_1 + \omega \alpha_2 + \omega^2 \alpha_3)
   )_{(q)}\;, \label{eq:pott-charged-moment}
 \end{align}
 where $ \alpha_i = \abs{a_i}^2$ and we use the basis  $ \{\chi_1(q), \chi_\varepsilon(q), \chi_\psi(q), \chi_\sigma(q), \chi_{\psi^\dagger}(q), \chi_{\sigma^\dagger}(q) \}$,    and
 \begin{align}
   \mu &= \sqrt{\frac{5 + \sqrt{5} }{30}} \;, \qquad \nu = - \sqrt{\frac{5 - \sqrt{5} }{30}} \;.
 \end{align}
 Note that $ Z_{\varepsilon \varepsilon}(q)$ only contains the characters $ \chi_1(q)$ and $ \chi_\varepsilon(q)$; therefore the final four rows in $ z_{\varepsilon \varepsilon}(q, \varepsilon)$ must be zero.\footnote{This is because, by construction, the Verlinde lines commute with the extended chiral algebra and therefore, by Schur's lemma, act as scalar multiples of the identity on irreducible representations of the chiral algebra \cite{petkovaGeneralizedTwistedPartition2001}.
 }
 This implies that $ \alpha_1 + \omega^2 \alpha_2 + \omega \alpha_3 =0$ and $ \alpha_1 + \omega \alpha_2 + \omega^2 \alpha_3 = 0$.
 Together, these constraints imply that $ \alpha_1 = \alpha_3$ and $ \alpha_2 = -\omega(\alpha_1 + \omega \alpha_3) = -\omega(1+\omega)\alpha_1$.
 Plugging these into the charged-moment \labelcref{eq:pott-charged-moment} yields
 \begin{align}
   z_{ \varepsilon \varepsilon}(q, \varepsilon) &= -(1+ \omega) \alpha_1\qty( \mu \chi_1(q) + \nu \chi_\varepsilon(q))\;.
 \end{align}
 Finally, we need a constraint to fix $ \alpha_1$.
 This is determined by the action of $ \mathcal{L}_\varepsilon$ on the boundary identity or $ \varepsilon$ conformal tower.
 At this stage, we do not know what that action is;
 for now, we will let $ \mathcal{L}_\varepsilon$'s action on the boundary identity conformal tower be some non-zero scalar $ \lambda$.
 This implies
 \begin{align}
   -(1+\omega) \alpha_1 \mu = \lambda \implies \alpha_1 = \lambda \sqrt{\frac{5 + \sqrt{5} }{30}}\;,
 \end{align}
 resulting in 
  \begin{align}
    z_{\varepsilon \varepsilon}(q, \varepsilon) &=   \lambda \chi_1(q) + \lambda (1 - \varphi) \chi_\varepsilon(q)\;. \label{eq:Potts-Charged-Moment}
  \end{align}
 As the coefficients in front  of $ \chi_1(q)$ and $ \chi_\varepsilon(q)$ are different in \cref{eq:Potts-Charged-Moment}, we conclude that the boundary identity conformal tower and the boundary $ \varepsilon$ conformal tower correspond to different irreps of $C$.
  As a consequence, for \cref{eq:Fib-projectors} to be projectors onto the irreducible representations of $C$ in the presence of boundaries, we must have $ \tr_{\varepsilon \varepsilon} \Pi_1^{\mathrm{bulk} } \rho_{A,\varepsilon \varepsilon} \stackrel{!}{=} \chi_1(q) $ or $ \chi_\varepsilon(q)$ and vice versa for $ \Pi_\varepsilon^{\mathrm{bulk} }$.
  
 Using the  charged moment $z_{\varepsilon \varepsilon}(q, \varepsilon)$ defined in \cref{eq:Potts-Charged-Moment} we compute 
 \begin{align}
   \tr_{\varepsilon \varepsilon} \Pi_1^{\mathrm{bulk} } \rho_{A,\varepsilon\varepsilon} 
   = \frac{ (1 + \lambda \varphi) \chi_1 + (1 - \lambda \varphi^2 + \lambda \varphi)\chi_\varepsilon }{1+\varphi^2}\;,
 \end{align}
 and assert
 \begin{align}
   \tr_{\varepsilon \varepsilon} \Pi_1^{\mathrm{bulk} } \rho_A \stackrel{!}{=} \chi_1 (q)\;. \label{eq:proj-assert}
 \end{align}
 However, no choice of $ \lambda$ satisfies the above, \cref{eq:proj-assert}.
 Similarly, no choice of $ \lambda$ can satisfy 
 \begin{align}
   \tr_{\varepsilon \varepsilon} \Pi_1^{\mathrm{bulk} } \rho_{A,\varepsilon\varepsilon} \stackrel{!}{=} \chi_\varepsilon(q)\;.
 \end{align}
 We conclude that the original bulk projectors \labelcref{eq:Fib-projectors} are incorrect in the presence of boundary conditions.
 
 The Fibonacci projectors in the bulk \labelcref{eq:Fib-projectors} leading to incorrect results in the presence of boundaries could be established on different, simpler grounds.
 Namely, the fusion algebra of defect lines terminating on a boundary is generally different than the bulk fusion algebra \cite{kojitaTopologicalDefectsOpen2018}.
 Following the prescription in Ref.~\cite{kojitaTopologicalDefectsOpen2018} we find that the boundary fusion rules for $ {C}  = \{1, \varepsilon\}$ with boundary condition $ \varepsilon$ on both sides is 
 \begin{align}
   \mathcal{L}_\varepsilon^{[\varepsilon]} \times \mathcal{L}_\varepsilon^{[\varepsilon]} &= \frac{1}{\varphi} \mathcal{L}_1^{[\varepsilon]} + \frac{\varphi - 1}{\varphi} \mathcal{L}_\varepsilon^{[\varepsilon]}\;,\label{eq:Potts-boundary-fusion}
 \end{align}
 where $ \mathcal{L}_c^{[\varepsilon]}$ indicates that the defect $ \mathcal{L}_c$ terminates on the two boundaries labelled by $\varepsilon$.
 From the above boundary fusion rules \labelcref{eq:Potts-boundary-fusion}, it is clear that the bulk projectors \labelcref{eq:Fib-projectors} do not satisfy $ \Pi_r^{\mathrm{bulk} } \Pi_h^{\mathrm{ bulk} } = \delta_{rh} \Pi_r^{\mathrm{bulk} } $ when the defects terminate on the boundary $ \varepsilon$.
 
 Before concluding this section, note that if one used the projectors 
 \begin{align}
   \Pi_1^{[\varepsilon]} &= \frac{1}{1+\varphi}(\mathcal{L}_1  + \varphi \mathcal{L}_\varepsilon)\;, \nonumber \\
   \Pi_\varepsilon^{[\varepsilon]} &= \frac{\varphi}{1+\varphi}(\mathcal{L}_1 - \mathcal{L}_\varepsilon)\;,  \label{eq:Potts-Fib-Projectors}
 \end{align}
 then for $ \lambda = 1$ we would find 
 \begin{align}
   \tr_{\varepsilon \varepsilon} \Pi_1^{[\varepsilon]} \rho_A &= \chi_1(q)\;, \nonumber \\
   \tr_{\varepsilon \varepsilon} \Pi_\varepsilon^{[\varepsilon]} \rho_A &= \chi_\varepsilon(q)\;.
 \end{align}
 We will show how to construct these projectors in \cref{sec:Constructing-Projectors}.
 
 \section{Topological defect networks}  \label{sec:TDL}
 In \cref{sec:Counter-Example}, the charged moments \labelcref{eq:charged-moments} were computed by taking overlaps of twisted boundary states.
 In Ref.~\cite{kusukiSymmetryresolvedEntanglementEntropy2023} the authors noted the similarity between computing these overlaps and orbifold CFTs, developing a method to compute the SREE by utilising boundary states in orbifold CFTs.
 The orbifold construction of the SREE makes it explicit why the symmetry of interest, $C$, needs to be anomaly free: constructing the orbifold CFT is only possible if $C$ is free of a 't Hooft anomaly.
 In contrast to Refs.~\cite{kusukiSymmetryresolvedEntanglementEntropy2023, saura-bastidaCategoricalsymmetryResolvedEntanglement2024}, we wish to compute the SREE strictly from the open string channel and only use the closed string channel to compute the breaking of equipartition in the asymptotic limit, \cref{eq:SREE-equipartition}.
 This requires knowledge of the action of defects on states in the open string Hilbert space without appealing to the closed string channel and twisted boundary states.
 
 When working in the open string channel, it is natural to use the graphical calculus afforded by fusion categories; 
 this requires performing manipulations of topological defect networks by utilising the appropriate $ F$-symbols.
 In the presence of boundaries, as is the case here, these manipulations can become difficult because one needs to keep track of the $ F$-symbols associated with the symmetry category of interest and the $ \tilde{F}$-symbols associated with the module category of interest (the boundary states).
 However, with simplifying assumptions, only a single set of $F$-symbols is required.
 We will focus on diagonal RCFTs in which charge conjugation acts trivially.
 As the CFT has a diagonal modular invariant, primary operators, Cardy boundary states, and Verlinde lines are in one to one correspondence and share the same labels in a modular fusion category $ \mathcal{M} $.
 Invariance under charge conjugation means that all Verlinde lines are self dual i.e. $ \mathcal{L}_c = \mathcal{L}_{\bar{c}}$ for $ c \in \mathcal{M} $. In other words, it is not necessary to specify the orientation of defects in diagrammatic manipulations. 
 For simplicity, we also restrict ourselves to situations where the fusion of Verlinde lines
 \begin{align}
   \mathcal{L}_c \times \mathcal{L}_d = \sum_{e}^{} N_{cd}^e \mathcal{L}_e
 \end{align}
    is multiplicity free. This means that the fusion coefficients $ N_{cd}^e$ are either $ 0$ or $ 1$ for all $c,d,e \in \mathcal{M}$. The fusion coefficients $N_{cd}^e$ can be determined from the modular S matrix with the Verlinde formula \cite{verlindeFusionRulesModular1988}.
    Finally, we assume parity invariance of our defects and CFTs.
 With these assumptions, we review the relevant formalism presented in Ref.~\cite{kojitaTopologicalDefectsOpen2018} to compute the action of defects on states in the open string Hilbert space.
 
 \subsection{Defect networks} 
 
 \begin{figure*}[]
   \centering
   \includegraphics{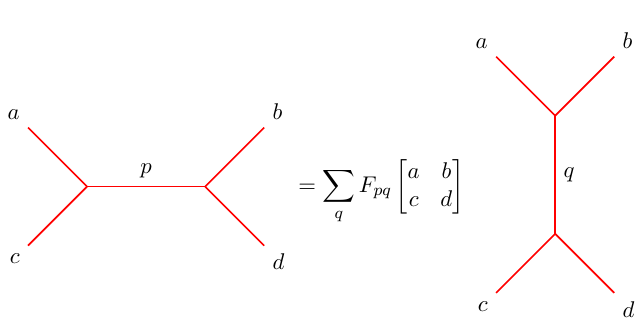}
   \caption{The $ F$-symbols allow manipulation of defect networks.}
   \label{fig:F-symbols}
 \end{figure*}
 Manipulations of networks of topological defects is done by utilising the $ F$-symbols.
 Our convention for the $ F$-symbols are shown in \cref{fig:F-symbols}.
 Moreover, if any of the incoming defects $ a$, $ b$, $  c$ or $ d$ are the identity defect, denoted by $ 1$, then the corresponding $ F$-symbol is trivial.
   For consistency of defect network manipulations, the $ F$-symbols must satisfy the  so-called pentagon identity; see, for instance, Ref.~\cite{kojitaTopologicalDefectsOpen2018}.
   This requires that the $F$-symbols satisfy
 \begin{align}
   \sum_{s}^{} F_{ps}\begin{bmatrix}
     b & c \\ a & q
   \end{bmatrix}
   F_{qt} \begin{bmatrix}
     s & d \\ a & e
   \end{bmatrix}
   F_{sr} \begin{bmatrix}
     c & d \\ b & t
   \end{bmatrix}
   \nonumber \\= F_{qr}\begin{bmatrix}
     c& d \\ p & e
   \end{bmatrix}
   F_{pt} \begin{bmatrix}
     b & r \\ a & e
   \end{bmatrix}\;. \label{eq:pentagon-identity}
 \end{align}
 
 As emphasised in Ref.~\cite{kojitaTopologicalDefectsOpen2018} (see also Ref.~\cite{Moore:1989vd}), the pentagon identity has a large symmetry, sometimes referred to as a gauge symmetry.
 From a set of $F$-symbols that satisfy the pentagon identity, applying an appropriate gauge transformation to these $F$-symbols results in another equivalent solution of the pentagon identity.
 Of course, after manipulating defect networks the final result obtained should be independent of the gauge choice (assuming the result corresponds to a physical observable).
 For our purposes, it is useful for our diagrammatic rules to be fully isotopy invariant; the defects can be freely moved and bent without incurring additional negative signs or phase factors.
 This condition corresponds to a specific choice of gauge for the $ F$-symbols which we will use in this work.
 In this gauge some of the $ F$-symbols simplify \cite{Simon:2023hdq} 
 \begin{align}
   F_{1c}\begin{bmatrix}
     a & b \\ a & b
   \end{bmatrix}= F_{c 1} \begin{bmatrix}
     a & a \\ b  & b
   \end{bmatrix} = \sqrt{\frac{d_c}{d_a d_b}} \;. \label{eq:isotopy-invariant}
 \end{align}
 Due to the assumption of parity invariance, we have $F$-symbol identities such as 
 \begin{align}
   F_{pq} \begin{bmatrix}
     a & b \\ c& d
   \end{bmatrix}
   = F_{pq} \begin{bmatrix}
     c &d \\ a & b
   \end{bmatrix}
   = F_{pq} \begin{bmatrix}
     b & a \\ d &c
   \end{bmatrix}\;.
 \end{align}
 
 \subsection{Open topological defects} 
\begin{figure*}[]
   \centering
   \includegraphics[]{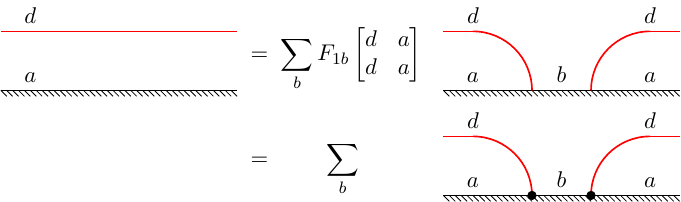}
   \caption{Fusing a defect parallel with a boundary naturally results in normalisation factors of the junction between the defect and boundary.
   These normalisation factors are denoted with a bold circle, as shown in the second line.}
   
   \label{fig:junction-norm}
 \end{figure*}

\begin{figure}[]
   \centering
   \includegraphics[width=\columnwidth]{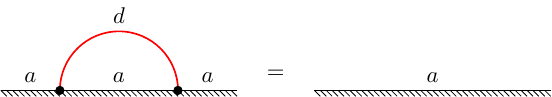}
   \caption{With the normalisation factors in \cref{fig:junction-norm}, defects can be shrunk on a boundary with no additional factors incurred.}
   \label{fig:defect-shrink}
 \end{figure}

 As explained in Ref.~\cite{kojitaTopologicalDefectsOpen2018}, attaching defects to a boundary comes with a choice of normalisation for the junction field.
 A natural choice of normalisation arises from using the $ F$-symbols to attach a parallel defect to a boundary and assigning a factor of $ \sqrt{F_{1b} {[}
     d  a; d a {]}} $ to each junction.
 This junction factor is denoted by a filled circle, see \cref{fig:junction-norm}.
 These junction factors have some convenient properties \cite{kojitaTopologicalDefectsOpen2018, konechnyOpenTopologicalDefects2020}.
 One particularly useful property we will utilise is the fact that, with these junction factors, we can shrink a defect on a  boundary to the identity, \cref{fig:defect-shrink}.
 
\begin{figure*}[]
   \centering
 \includegraphics[]{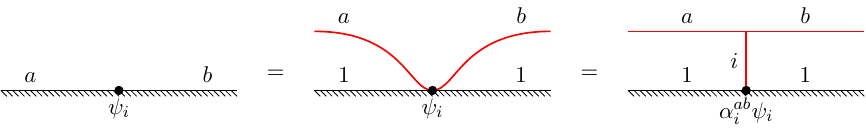}
   \caption{
   The boundary conditions $a$ and $b$ are separated from the reference, identity boundary $1$.
   The boundary condition changing operator $\psi_i$ is traded for a defect line with the same label $ i$, incurring a factor of $\alpha_i^{ab}$.}
   \label{fig:boundary-lift}
 \end{figure*}

 Next we require the action of defects on boundary condition changing operators.
 By the operator-state correspondence in CFT, boundary condition changing operators correspond to states in the open string Hilbert space \cite{cardyBoundaryConditionsFusion1989, cardyBoundaryConformalField2004}.
 Determining the action of a defect on boundary condition changing operators is then equivalent to determining the defect's action on states in the open string Hilbert space.
 Following Ref.~\cite{kojitaTopologicalDefectsOpen2018}, manipulations of boundaries with boundary condition changing operators can be lifted to defect manipulations by representing the boundary condition by the corresponding defect label.
 This assumes that there is some identity boundary that can be used to generate all other boundaries by fusion with an appropriate defect.
 In our case, this assumption is trivially satisfied by the vacuum Cardy state.
 Then, the boundary condition changing operator with representation label $ i$ can be traded for a defect labelled by $ i$ and a defect ending field on the identity boundary, \cref{fig:boundary-lift}.
 The constants $ \alpha_i^{ab}$ associated with this manipulation were computed in Ref.~\cite{kojitaTopologicalDefectsOpen2018}.
 However, the constants $ \alpha_i^{ab}$ will not be needed for our purposes.
 This manoeuvre, \cref{fig:boundary-lift}, also has an interpretation in terms of the topological field theory formulation of CFTs \cite{fuchsTFTConstructionRCFT2002,felderCorrelationFunctionsBoundary2002}.
 
 \subsection{Action of defects on boundary fields}  \label{sec:operator-state-action} 
 \begin{figure}[]
     \centering
 \includegraphics[width=\columnwidth]{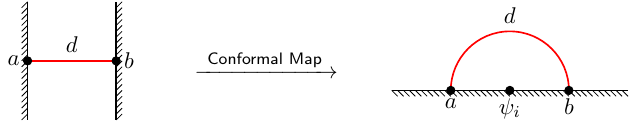}
     \caption{The annulus is mapped by a conformal transformation to the upper-half plane.
   By the operator-state correspondence, the action of a defect on the state $ \ket{i, M}$ in the open string Hilbert space, where $ i$ labels the boundary conformal tower and $ M$ enumerates the descendants, is equivalent to the action of the defect line on boundary condition changing operators (right).
 }
     \label{fig:annulus-to-UHP}
 \end{figure}
\begin{figure*}[]
   \centering
   \includegraphics[]{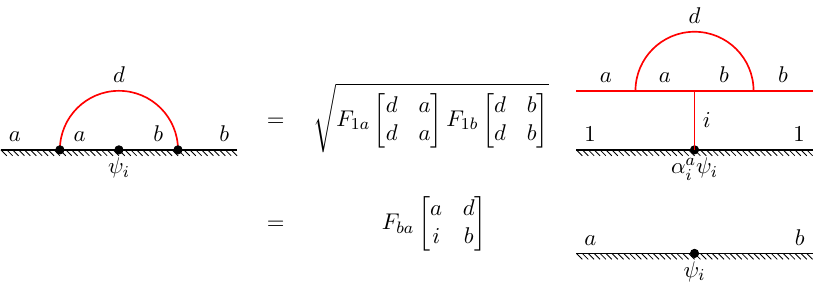} 
   \caption{The action of a defect $ \mathcal{L}_d$ on a boundary condition changing operator $ \psi_i$; 
   this utilises the manipulation shown in \cref{fig:boundary-lift}. }
   \label{fig:defect-action} 
 \end{figure*}

 
 To compute the SREE one needs to compute charged moments $ z_{a b}(q,d) = \tr_{a b} \mathcal{L}_d \rho_{A,ab}$.
 The insertion of the defect $ \mathcal{L}_d$ in this trace
 corresponds to connecting a defect to both boundaries, $ a$ and $ b$, of the annulus, \cref{fig:annulus-to-UHP}(left).
 By a conformal transformation, the annulus can be mapped to the upper-half plane and the defect then traverses a semi-circle around the origin, where a possible boundary condition changing operator is located, \cref{fig:annulus-to-UHP}(right).
 By the operator-state correspondence, a state $ \ket{i, M}$ in the open string Hilbert space,  where $ i$ labels a boundary conformal tower and $ M$ enumerates its states, is created by a local boundary condition changing operator $ \psi_{i, M}$ \cite{cardyBoundaryConditionsFusion1989}.
 The action of the defect operator $ \mathcal{L}_d$ on states in the open string Hilbert space is determined by the action of $ \mathcal{L}_d$ on boundary condition changing operators.
 With the assumptions specified at the start of \cref{sec:TDL} -- primary operators, boundary states and defect lines are in one-to-one correspondence -- this action has been derived in Ref.~\cite{kojitaTopologicalDefectsOpen2018}.
 This action is illustrated in \cref{fig:defect-action} under the assumption that the boundaries at the entangling surface are described by simple boundary conditions.
 For the generalisation to non-simple boundary conditions see Ref.~\cite{kojitaTopologicalDefectsOpen2018}.
 If the boundary conditions $ a$ and $ b$ are not simple then the action of defect operators is more complicated.
 This is because the defect operator can now intertwine different boundary condition changing operators in the same representation.
 
 In the BCFT approach to the SREE, \cref{sec:SREE-BCFT-review}, we required our boundary states to be weakly symmetric under our fusion category of interest.
 We now see how this requirement manifests itself in the open string Hilbert space.
 The requirement of the defect $ d$ being able to topologically end on $ a$ and $ b$ is encoded in the $ F$-symbol $ F_{ba}[ ad; ib]$, see \cref{fig:defect-action}, which specifies the action of $ \mathcal{L}_d$ on states in the open string Hilbert space.
 If $ a$ and $ b$ are not weakly symmetric under $ d$ then $ F_{ba}[ad; ib] = 0$ for all $ i$, implying that $ \mathcal{L}_d = 0 $ in the open string Hilbert space.
 
\begin{figure*}[]
   \centering
   \includegraphics[]{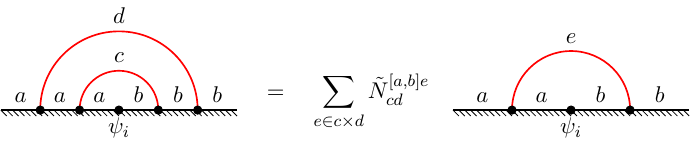}
   \caption{Fusion of defects in the presence of boundaries.}
   \label{fig:boundary-fusion}
 \end{figure*}

 The fusion of two defects in the presence of boundaries $ a$ and $ b$ is denoted by
 \begin{align}
   \mathcal{L}_c^{[a,b]} \times \mathcal{L}_d^{[a,b]} = \sum_{ e \in c \times d}^{} \tilde{N}_{cd}^{[a,b] e} \mathcal{L}_e^{[a,b]}\;, \label{eq:boundary-fusion}
 \end{align}
 where $ \tilde{N}_{cd}^{[a,b] e}$ are generally non-integer coefficients \cite{kojitaTopologicalDefectsOpen2018}.
 By using the $ F$-symbols to fuse two defects in the presence of boundaries, as in \cref{fig:boundary-fusion}, and utilising \cref{eq:isotopy-invariant}, the boundary fusion coefficients are given by
 \begin{align}
   \tilde{N}_{cd}^{[a,b]e} = F_{eb} \begin{bmatrix}
     d & c \\ b & b
   \end{bmatrix}
   F_{ea} \begin{bmatrix}
     d & c \\ a & a
   \end{bmatrix}\;. \label{eq:boundary-fusion-rules}
 \end{align}
 Peculiarly, even if $e$ appears in the bulk fusion $ c \times d$, the boundary fusion coefficient $\tilde{N}_{cd}^{[a,b] e}$ \labelcref{eq:boundary-fusion-rules} may be zero, resulting in $\mathcal{L}_e$ not appearing on the right hand side of \cref{eq:boundary-fusion}.
 We will see an explicit example of this when we study Rep($S_3$) in the presence of boundaries in \cref{sec:Rep(S3)}.
 With our choice of junction normalisation, and specialising to the case $ a = b$,  we note some identities of the boundary fusion rules \cite{konechnyOpenTopologicalDefects2020}:
 \begin{align}
   \tilde{N}_{1d}^{[a]d} = 1\;, \qquad \tilde{N}_{dd}^{[a] 1} = \frac{1}{d_d}\;, \qquad \sum_{e \in c \times d}^{} \tilde{N}_{cd}^{[a] e} = 1\;.  \label{eq:boundary-fusion-identities}
 \end{align}
 The last identity arises due to the fact that defects can be shrunk on a boundary with no additional factors incurred, \cref{fig:defect-shrink}.
 
 \section{Boundary projectors}  \label{sec:Constructing-Projectors}
 
 \subsection{Projector construction}
 
 We are now in a position to use Verlinde lines to project onto sectors of the open string Hilbert space.
 First, we will construct projectors that project onto the boundary conformal towers in the open string Hilbert space.
 Before we proceed, recall that all boundary states, Verlinde lines and primary operators share labels in the modular fusion category $ \mathcal{M} $ and the fusion of defects is multiplicity free.
 
 The open string Hilbert space decomposes as $ \mathcal{H}_{ab} = \bigoplus_{r} N_{ab}^r \mathcal{V}_r $.
 We want to find operators $ P_r^{[a,b]}$ that implement the projections $ P_r^{[a,b]}: \mathcal{H}_{ab} \rightarrow N_{ab}^r \mathcal{V}_r $, i.e.
 \begin{align}
   \tr_{ab} \Pi_r^{[a,b]} \rho_{A,ab} = N_{ab}^r\chi_r(q)\;.
 \end{align}
 These projectors were constructed on a formal level in Ref.~\cite{northeEntanglementResolutionRespect2023},
 \begin{align}
   P_r = \sum_{M}^{} \ket{r,M}\bra{r,M}\;, \label{eq:projectors-formal}
 \end{align}
 where $ M$ is a descendant label.
 Ref. \cite{northeEntanglementResolutionRespect2023} used \cref{eq:projectors-formal} to analyse the SREE with respect to conformal symmetry.
 
 Firstly, let $ \mathrm{WS}^{[a,b]} \subseteq \mathcal{M}$ denote the subset of Verlindes lines that both boundaries $ a$ and $ b$ are weakly symmetric under.\footnote{\JHc{For $a = b$ we have $\mathrm{WS}^{[a,b]} = a \times b$.
 However, if $a \neq b$ then $\mathrm{WS}^{[a,b]} \neq a \times b$. This is easily seen by observing that the identity label is always in $\mathrm{WS}^{[a,b]}$, but only in $a \times b$ when $ a = b$ (under the assumption of self-dual defects).}}
 We propose that the projector onto $ \mathcal{V}_r$ is 
 \begin{align} 
   P_r^{[a,b]} &= \frac{1}{F_{1b}\mqty[a & r \\a &r ]} \sum_{s\in \mathrm{WS}^{[a,b]}}^{} F_{1s}\mqty[a & a \\\ a & a] F_{sr}\mqty[a & b \\ a & b] \mathcal{L}_s \label{eq:BCFT-projector} \;.
 \end{align}
 By using the following $ F$-symbol identity, which can be derived from the pentagon identity (see \cref{sec:pentagon-id-proj})
 \begin{align}
   \frac{1}{F_{1b} \begin{bmatrix}
       a & r \\ a & r
   \end{bmatrix}} \sum_{s \in \mathrm{WS}^{[a,b]}}^{} F_{1s} \begin{bmatrix}
       a & a \\ a & a
   \end{bmatrix} F_{sr} \begin{bmatrix}
       a & b \\ a & b
   \end{bmatrix}F_{b a} \begin{bmatrix}
       a & s \\ i & b
     \end{bmatrix}  = \delta_{i r}\;, \label{eq:pentagon-special}
   \end{align}
 we have
 \begin{align}
   \tr_{ab} P_r^{[a,b]} \rho_{A,ab} &= \sum_{i}^{} N_{ab}^i \frac{1}{F_{1b} \begin{bmatrix}
       a & r \\ a & r
   \end{bmatrix}}\nonumber \sum_{s \in \mathrm{WS}^{[a,b]}}^{} F_{1s} \begin{bmatrix}
       a & a \\ a & a
   \end{bmatrix} 
   \nonumber \\ 
   &\qquad \times F_{sr} \begin{bmatrix}
       a & b \\ a & b
   \end{bmatrix}F_{b a} \begin{bmatrix}
       a & s \\ i & b
   \end{bmatrix}  \chi_i(q) \nonumber \\
   &= N_{ab}^r \chi_r(q)\;,
 \end{align}
 as required.
 Using \cref{eq:pentagon-special}, it is easy to show that $ \{P_r^{[a,b]}\}_{r\in a \times b}$ are orthogonal projectors when acting on states in the open string Hilbert space,
 \begin{align}
   P_h^{[a,b]}P_r^{[a,b]} \ket{i,M}= \delta_{rh} P_r^{[a,b]} \ket{i,M}\;.
 \end{align}
 Similarly, using \cref{eq:isotopy-invariant} and the pentagon identity \eqref{eq:pentagon-identity}, one can show
 \begin{align}
     \sum_{\JHc{r\in a \times b}} P_r^{[a,b]} = \mathcal{L}_1\;,\label{eq:projector-complete}
 \end{align}
 \JHc{showing that the projectors are complete.}
 
 \subsection{Asymptotic limit}  \label{sec:asymptotic-limit}
 Utilising the projectors \eqref{eq:BCFT-projector}, the asymptotic limit of the SREE can be computed.
 In the asymptotic limit, the untwisted sector in the closed string channel dominates. 
 Therefore 
 \begin{align}
   \mathcal{Z}_{ab}(q,r) &=  \tr_{ab} P_r^{[a,b]} \rho_{A,ab} \nonumber \\ &\stackrel{q\to1}{\sim} \frac{F_{11} \begin{bmatrix}
       a & a \\ a & a
   \end{bmatrix} F_{1 r} \begin{bmatrix}
       a & b \\ a & b
   \end{bmatrix}}{F_{1b} \begin{bmatrix}
       a & r \\ a & r
   \end{bmatrix}} \bra{a}\ket{0}\bra{0}\ket{b} \tilde{q}^{c/24} \;, \nonumber \\
      &= \frac{d_r}{d_a d_b} \bra{a}\ket{0}\bra{0}\ket{b} \tilde{q}^{c/24}\;, \nonumber \\
     &= S_{1r} \tilde{q}^{c/24}\;,   \label{eq:asymptotic-1}
 \end{align}
 where we have explicitly expanded the Cardy states in terms of the modular $ S$ matrix to go from the second last to the final line.
 Unsurprisingly, the final, simplified result is identical to results found in Ref.~\cite{northeEntanglementResolutionRespect2023}, which utilised the projectors in \cref{eq:projectors-formal}.
 This serves as a useful check of the validity of the BCFT projectors in \cref{eq:BCFT-projector}.
 Using \cref{eq:SREE-equipartition}, we find that the equipartition of the SREE is broken by the term $ \log d_r/d_a d_b$ when projecting onto the conformal family  $ r$.
 
 Currently, the projectors \cref{eq:BCFT-projector} do not provide any insight on the asymptotic limit of the SREE for a fusion category symmetry because it is not obvious how the coefficients $d_r/d_a d_b$ in \cref{eq:asymptotic-1} relate to \JHc{the symmetry defects} $\mathrm{WS}^{[a,b]}$.
 More insight can be gained with some additional assumptions.
 \JHc{Namely, we fix both boundaries to be identical, $a =b = a_C$, where $a_C$ is chosen such that it is symmetric under a symmetry category $C$ and $a_C \times a_C = \sum_{c \in C} c$.}
 \footnote{\JHc{The open string partition function will be \begin{align}
     Z_{a_C a_C} = \sum_{r \in C} \chi_r(q)\;.
 \end{align}}}
 \JHc{
 In this setup, the set of Verlinde lines that $a_C$ is weakly symmetric under is $\mathrm{WS}^{[a_C]} = C $.}
 As the quantum dimensions furnish a representation of the fusion ring, it follows that
 \begin{align}
   \sqrt{C} \coloneq d_{a_C}^2 &= \sum_{c \in C }^{} d_c\;.
 \end{align}
 Using \cref{eq:BCFT-projector}, we construct the projectors onto the irreps of $C$
 \begin{align}
   P_r^{[a_C]} &= \frac{1}{F_{1a_C} \begin{bmatrix}
       a_C & r \\ a_C & r
   \end{bmatrix}}
   \sum_{d \in C }^{} F_{1d} \begin{bmatrix}
     a_C & a_C \\ a_C & a_C
   \end{bmatrix}
   F_{dr} \begin{bmatrix}
      a_C& a_C \\ a_C & a_C
   \end{bmatrix}
   \mathcal{L}_d\;,
 \end{align}
 which can be written in the nicer form
 \begin{align}
   P_r^{[a_C]}  = \frac{d_r}{\sqrt{C} } \sum_{d \in C}^{} d_d  F_{a_C a_C} \begin{bmatrix}
     d & a _C \\ a_C & r
   \end{bmatrix}
   \mathcal{L}_d\;. \label{eq:MFC-general?}
 \end{align}
 The boundary fusion rules \eqref{eq:boundary-fusion-rules} can be used to explicitly show that $ \{P_r^{[a_C]}\}_{r \in C}$ are orthogonal projectors,
 \begin{align}
   P_r^{[a_C]} P_h^{[a_C]} &= \delta_{rh} P_r^{[a_C]}\;. \label{eq:orthogonal-projectors}
 \end{align}
 This is a tedious calculation, so we differ it to \cref{sec:appendix-projector-calc}.
 Additionally, from \cref{eq:projector-complete} we have
 \begin{align}
     \sum_{r \in C} P_r^{[a_C]} = \mathcal{L}_1.
 \end{align}
 
 Using \cref{eq:SREE-equipartition}, we see that the equipartition of the SREE is broken by the term
 \begin{align}
    \Delta S_n(q,r) = \log \frac{d_r}{\sqrt{{C} } }\;. \label{eq:MFC-SREE-equipartition}
 \end{align}
 This is in contrast with the results obtained in Ref.~\cite{saura-bastidaCategoricalsymmetryResolvedEntanglement2024}, \cref{eq:equipartition-wrong}, which suggest that the equipartition should be broken by $ \log d_r^2/\abs{{C} }^2 $ where $ \abs{C}^2  = \sum_{c \in C}^{} d_c^2$.
 \cref{eq:MFC-SREE-equipartition} is in agreement with Ref.~\cite{kusukiSymmetryresolvedEntanglementEntropy2023}, \cref{eq:equipartition-group}, when $ {C} $ is a finite abelian group 
 because irreducible representations of finite abelian groups are one dimensional.
 Strictly speaking, due to the assumption of self-duality of representations, $C$ must be a product of $\mathbb{Z}_2$-factors.
 However, we believe this assumption can easily by relaxed without changing the outcome, \cref{eq:MFC-SREE-equipartition}.
 While \cref{eq:MFC-SREE-equipartition} disagrees with the results of Ref.~\cite{kusukiSymmetryresolvedEntanglementEntropy2023}, \cref{eq:equipartition-group}, for non-abelian finite groups, this disagreement is superficial: $ {C} $ can never be a non-abelian group because the fusion of Verlinde lines is commutative.
 
 A natural concern one may have is the generality of the projectors, \cref{eq:MFC-general?}.
 Indeed, the construction of \cref{eq:MFC-general?} required a careful choice of boundary conditions.
 The boundary conditions are referenced explicitly in the boundary fusion rules \eqref{eq:boundary-fusion-rules} so it would appear that the projectors constructed in \cref{eq:MFC-general?} and all computations based on them are true specifically for the chosen boundary condition $ a_C$.
 This is a valid concern, and we cannot definitively establish at the time being that \cref{eq:MFC-general?} are the general projectors that project onto irreducible representations of $ C$ in the presence of more general boundary conditions.
  Currently, \cref{eq:MFC-general?} can be used on a case-by-case basis.
 Suppose that one is studying a theory with boundary conditions $ a'$ and $ b'$ that are weakly symmetric under $ C$.
 Due to the orthogonality condition \labelcref{eq:orthogonal-projectors}, a sufficient condition for the projectors with boundaries $a_C$ \labelcref{eq:MFC-general?} to be projectors in the presence of boundaries $a'$ and $b'$ is requiring that the fusion of defects in $C$ does not depend on whether the boundaries are labelled by $a_C$ or $a'$ and $b'$:
 \begin{align}
   \tilde{N}_{dc}^{[a_C] e} = \tilde{N}_{dc}^{[a',b'] e}\;, \label{eq:fusion-equal}
 \end{align}
 for all $ d,c,e \in C$.
 If \cref{eq:fusion-equal} is true, then \cref{eq:MFC-general?} will project onto the irreps of $ C$ in the theory with boundaries $ a'$ and $ b'$ even though they were constructed with the boundary $a_C$.
 We will show an example of this when we study the tricritical Ising model in \cref{sec:Tricritical-Ising}.
 
 Given the explicit dependence of the boundary fusion rules on the choice of boundary conditions, \cref{eq:boundary-fusion-rules}, it would be quite surprising if \cref{eq:fusion-equal} turned out to be true in full generality. 
 In the next section, after studying explicit examples of the boundary projectors \eqref{eq:MFC-general?}, we will briefly discuss the validity of \cref{eq:fusion-equal} in the context of $ C = \mathrm{Fib} $ and $ C = \mathrm{Rep}(S_3) $ from the perspective of anyonic chains.
 
 Before we study examples of the BCFT projectors, \cref{eq:BCFT-projector} and \cref{eq:MFC-general?}, we note that it is impossible to construct the projectors onto the irreducible representations of $C$ \eqref{eq:MFC-general?} by only using the $ F$-symbols of $ C$ and making no reference to a boundary condition.\footnote{We restrict the data to only the $ F$-symbols of $ C$ because $ C$ need not be a modular fusion category.} 
 We will comment on this further when we study the Ising model in \cref{sec:Ising}.
 It may be possible to write the projectors \labelcref{eq:MFC-general?} in terms of other data of $ C$, such as the characters of $ C$ in the presence of boundaries.
 The boundary characters $\tilde{\chi}_i^{[a,b]}(c)$ should form a one-dimensional representation of the boundary fusion algebra
 \begin{align}
     \tilde{\chi}_i^{[a,b]}(c) \tilde{\chi}_j^{[a,b]}(c) = \sum_{k \in i \times j} \tilde{N}_{ij}^{[a,b] k} \tilde{\chi}_k(c)^{[a,b]}\;,
 \end{align}
 where $c \in C$ and irreps of $C$ are labelled by  $i,j,k \in C$.
 Moreover, they should obey relevant orthogonality relations and reduce to suitable (quantum) dimensions under appropriate circumstances, analogous to the finite group characters, to befit being called a character.\footnote{In the bulk the natural characters $\chi_j(i)=S_{ij}/S_{i1}$ would be quotients of modular S-matrix entries.}
 Again, it seems likely that one would also have to make an explicit reference to boundary conditions to define the characters $\tilde{\chi}_i^{[a,b]}(c)$ because defining the boundary fusion rules rules requires a choice of boundaries. 
 Therefore, from a mathematical perspective, it is unclear whether formulas for fusion category projectors in the presence of boundaries can be constructed without explicit reference to the boundaries.
 Despite this, in \cref{sec:Boundary-Independence} we appeal to physical arguments to suggest that such projectors are independent of the choice of boundary conditions.
 
 \section{Examples}  \label{sec:MFC-SREE-examples}
 
   After our general discussion of the theoretical underpinnings we now apply the above framework to concrete examples. The $F$-symbols used in this section have been obtained utilising Refs.~\cite{vercleyenLowRankFusion2023, ardonneClebschGordan6jcoefficientsRank2010}.
 
 \subsection{Revisiting the three-state Potts model.}
 
 Our goal is to find the projectors \eqref{eq:Potts-Fib-Projectors} using \cref{eq:BCFT-projector}. The action of $ \mathcal{L}_1$ on the boundary conformal towers  $ 1$ and $ \varepsilon$ is trivial.
 The action of $ \mathcal{L} _\varepsilon$ is trivial on the boundary conformal tower  $ 1$, while the action of $ \mathcal{L}_\varepsilon$  on the boundary conformal tower  $ \varepsilon$ is given by
 \begin{align}
   F_{\varepsilon\varepsilon}\begin{bmatrix} 
     \varepsilon & \varepsilon \\ \varepsilon & \varepsilon
   \end{bmatrix}\; = 1- \varphi\;.
 \end{align}
 This is in agreement with taking $ \lambda = 1$ in \cref{eq:Potts-Charged-Moment}.
 Finally, using the boundary fusion rules \labelcref{eq:boundary-fusion-rules}, the non-trivial fusion rules are\footnote{We note that boundary fusion rules for the Fibonacci category appear in Refs.~\cite{choiRemarksBoundariesAnomalies2023, kirchnerCharacterizingEntanglementAnyonic2024}, however, their choice of junction normalisation differs from ours.}
 \begin{align}
   \mathcal{L}_\varepsilon^{[\varepsilon]} \times \mathcal{L}_\varepsilon^{[\varepsilon]} = \frac{1}{\varphi} \mathcal{L}_1^{[\varepsilon]} + \frac{\varphi -1}{\varphi} \mathcal{L}_\varepsilon^{[\varepsilon]}\;. \label{eq:W-fusion}
 \end{align}
 Using these fusion relations, it can be checked that the Fibonacci projectors in the three-state Potts model \eqref{eq:Potts-Fib-Projectors} are orthogonal.
 
 \subsection{$\mathbb{Z}_2$: Ising Model} \label{sec:Ising}
 
 The Ising model has a $ \mathbb{Z}_2$ symmetry generated by $\varepsilon$; this $ \mathbb{Z}_2$ fusion category is a subcategory of the Ising modular fusion category and can thus be used within our formalism.
 The Cardy state $ \ket{\sigma}$, corresponding to the primary spin operator, is weakly symmetric with respect to the $ \mathbb{Z}_2$ symmetry.
 The open string partition function with boundaries $ \sigma$ is 
 \begin{align}
   Z_{\sigma\sigma}(q) = \tr_{\sigma \sigma}\rho_{A,\sigma\sigma} = \chi_1(q) + \chi_\varepsilon(q)\;.
 \end{align}
 From Ref.~\cite{kusukiSymmetryresolvedEntanglementEntropy2023}, we know that $ \chi_1(q)$ is the $ \mathbb{Z}_2$ even sector and $ \chi_\varepsilon(q)$ is the $ \mathbb{Z}_2$ odd sector.
 Using \cref{eq:BCFT-projector}, we find the following projectors
 \begin{align}
   P_1^{[\sigma]} &= \frac{1}{2}\qty(\mathcal{L}_1 + \mathcal{L}_\eta)\;, \nonumber \\
   P_\varepsilon^{[\sigma]} &= \frac{1}{2}\qty(\mathcal{L}_1 - \mathcal{L}_\varepsilon)\;, \label{eq:Ising-projectors}
 \end{align}
 which are nothing other than $ \mathbb{Z}_2$ projectors!
 The action of $ \mathcal{L}_\varepsilon$ on the $ \varepsilon$ boundary conformal tower is given by the $ F$-symbol
 \begin{align}
   F_{\sigma \sigma} \mqty[ \sigma & \varepsilon \\ \varepsilon & \sigma] = -1\;.
 \end{align}
 From this, it follows that
 \begin{align}
   \tr_{\sigma \sigma} P_1^{[\sigma]} \rho_{A,\sigma \sigma} = \chi_1(q)\;, \nonumber \\
   \tr_{\sigma \sigma} P_\varepsilon^{[\sigma]} \rho_{A, \sigma\sigma} = \chi_\varepsilon(q)\;,
 \end{align}
 in agreement with \cite{kusukiSymmetryresolvedEntanglementEntropy2023}.
 
 Finally, note that the $ \mathbb{Z}_2$ projectors \eqref{eq:Ising-projectors} cannot be constructed in terms of the $ F$-symbols of the (non-anomalous)\footnote{For a category $C$ to kinematically allow a (weakly) symmetric boundary state, it needs to be anomaly free \cite{choiRemarksBoundariesAnomalies2023}.} $ \mathbb{Z}_2$ fusion category.
 All the $ F$-symbols in $ \mathbb{Z}_2$ are trivial, so the coefficient of $ -1/2$ in \cref{eq:Ising-projectors} could not possibly be expressed in terms of these $F$-symbols. 
 To obtain this negative sign with the $F$-symbols, we need to make reference to a boundary condition, in this case $ \sigma$.
 
 \subsection{Tricritical Ising Model} \label{sec:Tricritical-Ising}
 To make contact with the original Cat-SREE paper \cite{saura-bastidaCategoricalsymmetryResolvedEntanglement2024}, we will analyse the tricritical Ising model in greater depth.
 Following conventions in the literature \cite{choiRemarksBoundariesAnomalies2023}, the primary operators are labelled as 
 \begin{align}
   {1}\;, \quad \eta\;, \quad \mathcal{N}\;, \quad W\;, \quad \eta W\;, \quad \mathcal{N} W\;.
 \end{align}
 The tricritical Ising model is a diagonal RCFT: there is a one-to-one correspondence between boundary states and Verlinde lines.
 One finds that the three (Cardy) boundary states $ \ket{W}$, $ \ket{\eta W} $ and $ \ket{\mathcal{N} W}$ are weakly symmetric under \JHc{the Fibonacci category} $C \JHc{= \{1, W \}}$:
 \begin{align}
   W \times W= W + 1\;, \qquad W \times \eta W &= \eta W + \eta\;, \nonumber \\ W \times  \mathcal{N} W &=  \mathcal{N} W + \mathcal{N} \; .
 \end{align}
 In the basis  $ \{{1}, \eta W, W, \eta, \mathcal{N}  W, \mathcal{ N} \}$ 
 the modular $ S$ matrix is given by \cite{difrancescoConformalFieldTheory1997}
 \begin{align}
   S=\left(\begin{array}{cccccc}
 s_2 & s_1 & s_1 & s_2 & \sqrt{2} s_1 & \sqrt{2} s_2 \\
 s_1 & -s_2 & -s_2 & s_1 & \sqrt{2} s_2 & -\sqrt{2} s_1 \\
 s_1 & -s_2 & -s_2 & s_1 & -\sqrt{2} s_2 & \sqrt{2} s_1 \\
 s_2 & s_1 & s_1 & s_2 & -\sqrt{2} s_1 & -\sqrt{2} s_2 \\
 \sqrt{2} s_1 & \sqrt{2} s_2 & -\sqrt{2} s_2 & -\sqrt{2} s_1 & 0 & 0 \\
 \sqrt{2} s_2 & -\sqrt{2} s_1 & \sqrt{2} s_1 & -\sqrt{2} s_2 & 0 & 0
 \end{array}\right)
 \end{align}
 with
 \begin{align}
   s_1 = \sin(2\pi /5) \qquad s_2 = \sin(4\pi /5)\;. 
 \end{align}
 The Verlinde lines labelled by $ C = \{ {1}, W \}$ form a Fibonacci subcategory; we with to resolve the entanglement entropy with respect to $C$.
 To compute the SREE, Ref.\ \cite{saura-bastidaCategoricalsymmetryResolvedEntanglement2024} considered the case when the boundaries at the entangling surface were $a = b =  {\mathcal{N} W}$.
 With these boundaries, the open string partition function is 
 \begin{align}
   Z_{\mathcal{N}W | \mathcal{N}W}(q) = \chi_1(q) + \chi_{\eta W}(q) + \chi_W(q) + \chi_{\eta}(q)\;.
 \end{align}
 To construct projectors onto the irreps of $C$, we first need to determine how $C$ acts on each boundary conformal tower, \cref{fig:defect-action}; we require the $ F$-symbols
 \begin{align}
   F_{\mathcal{N}W| \mathcal{N}W}\mqty[ \mathcal{N}W & d \\ i & \mathcal{N}W]\;,
 \end{align}
 with $ d \in C$ and  $ i \in \{1, \eta W, W, \eta \}$.
 If $ d$ or $ i$ is $ 1$ then this action is simply the identity.
 Therefore, the relevant $ F$-symbols are
 \begin{align}
   F_{\mathcal{N}W| \mathcal{N}W}\mqty[ \mathcal{N}W & W \\ W & \mathcal{N}W ] &=  1 -\varphi\;, \nonumber \\
 F_{\mathcal{N}W| \mathcal{N}W}\mqty[ \mathcal{N}W & \eta W \\ W & \mathcal{N}W ]  &= 1 -\varphi\;,\quad  \nonumber \\
 F_{\mathcal{N}W| \mathcal{N}W}\mqty[ \mathcal{N}W & \eta  \\ W & \mathcal{N}W ]  &= 1\;. \label{eq:Tricrit-F-symbols}
 \end{align}
 We conclude that the $ 1$ and $ \eta$ boundary conformal towers transform trivially under $ C$, while the $ \eta W$ and $ W$ boundary conformal towers transform non-trivial under $ C$.
 The projectors onto the trivial, $ 1$, and non-trivial, $ W$, irreducible representation of $ C$ are then given by
 \begin{align}
   \Pi_1^{[\mathcal{N}W]} &= P_1^{[\mathcal{N} W]} + P_\eta^{[\mathcal{N}W]}\;, \nonumber \\
 \Pi_W^{[\mathcal{N}W]} &= P_W^{[\mathcal{N} W]} + P_{\eta W}^{[\mathcal{N}W]}\;.
 \end{align}
 Using \cref{eq:BCFT-projector} we find 
 \begin{align}
   \Pi_1^{[\mathcal{N}W]} &= \frac{1}{1+\varphi}(\mathcal{L}_1 + \varphi \mathcal{L}_W)\;, \nonumber \\
   \Pi_W^{[\mathcal{N}W]} &= \frac{\varphi}{1+\varphi}(\mathcal{L}_1 - \mathcal{L}_W)\;. \label{eq:Tricrit-Project}
 \end{align}
 Using the action of $C$ on each boundary conformal tower \labelcref{eq:Tricrit-F-symbols}, one can verify that the above projectors \labelcref{eq:Tricrit-Project} implement 
 \begin{align}
   \tr_{\mathcal{N}W|\mathcal{N}W} \Pi_1^{[\mathcal{N} W]} \rho_{A,\mathcal{N}W|\mathcal{N}W} = \chi_1(q) + \chi_\eta(q)\;, \nonumber \\
   \tr_{\mathcal{N}W|\mathcal{N}W} \Pi_W^{[\mathcal{N} W]} \rho_{A,\mathcal{N}W|\mathcal{N}W} = \chi_{\eta W}(q) + \chi_{W}(q)\;.
 \end{align}
 Note that the projectors we found, \cref{eq:Tricrit-Project}, are identical to those found in the three state Potts model, \cref{eq:Potts-Fib-Projectors}.
 Indeed, as both boundaries are identical, $ a = b = \mathcal{N} W$, the boundary fusion rule identities \labelcref{eq:boundary-fusion-identities} can be used to show the non-trivial boundary fusion rules of $C$ are identical to those of the three-state Potts model:
 \begin{align}
   \mathcal{L}_W^{[\mathcal{N} W]} \times \mathcal{L}_W^{[\mathcal{N} W]} = \frac{1}{\varphi} \mathcal{L}_1^{[\mathcal{N} W]} + \frac{\varphi -1}{\varphi } \mathcal{L}_W^{[\mathcal{N } W]}\;.    
 \end{align}
 The breaking of the equipartition of the SREE \labelcref{eq:SREE-equipartition} is then
 \begin{align}
     \Delta S_n(q, 1) = \log \frac{1}{1+ \varphi}\;, \qquad \Delta S_n(q, W) = \log \frac{\varphi}{1+\varphi}\;.
 \end{align}
 This is in contradiction with Ref.~\cite{saura-bastidaCategoricalsymmetryResolvedEntanglement2024}, which suggest that $\Delta S_n(q,1)$ and $\Delta S_n(q,W)$ are $\log 1/(1+\varphi^2)$ and $\log \varphi/(1+\varphi^2)$, respectively.
 
 Next, let us consider the case of two different boundaries, $a = \eta W$ and $b = \mathcal{N}W$.
 The open string partition function is 
 \begin{align}
     Z_{\eta W| \mathcal{N}W}(q) = \chi_{\mathcal{N}}(q) + \chi_{\mathcal{N}W}(q)\;.
 \end{align}
 Before we appeal to the projector formula \labelcref{eq:BCFT-projector}, let us illustrate the logic surrounding \cref{eq:fusion-equal}.
 In the previous example, we have constructed the projectors onto the irreps of $C$ when both boundaries are $\mathcal{N} W$.
 These projectors can be used with the new boundary conditions, $a = \eta W$ and $b = \mathcal{N}W$, if the boundary fusion rules of $C$ are unchanged with these new boundaries.
 Using \cref{eq:boundary-fusion-rules} we find
 \begin{align}
     \tilde{N}_{dc}^{[\mathcal{N} W] e} =& \tilde{N}_{dc}^{[\eta W, \mathcal{N}W]e } \;,
 \end{align}
 for $d$, $c$, $e \in C$.
 Therefore, the projectors onto the irreps of $C$ with boundaries $a = \eta W$ and $b = \mathcal{N} W$ are also given by \cref{eq:Tricrit-Project}.
 
 To know which boundary conformal towers transform in the trivial and non-trivial irreps of $C$, we can: appeal to explicit $F$-symbols, as we did in \cref{eq:Tricrit-F-symbols},  construct the projectors onto each conformal tower with \cref{eq:BCFT-projector}, or analyse the asymptotic SREE by computing \labelcref{eq:asymptotic-1} $d_r/d_{\eta W} d_{\mathcal{N}W }$ for $r \in \eta W \times \mathcal{N} W$.
 Using the latter option, we find $d_\mathcal{N}/d_{\eta W} d_{\mathcal{N}W } = 1/(1 + \varphi) $ and $d_r/d_{\eta W} d_{\mathcal{N}W } = \varphi/(1+\varphi)$,
 indicating that the $\mathcal{N}$ and $\mathcal{N}W$ boundary conformal towers transform in the trivial and non-trivial irrep of $C$, respectively.
 
 \subsubsection{Numerical check: general remarks}\label{sec:num-general}

 \TQc{In this and the following subsection, we simulate two critical anyonic chains, the golden chain \cite{feiguinInteractingAnyonsTopological2007} with a Fibonacci category symmetry and a system with $\rm{Rep}(S_3)$ symmetry \cite{gilsAnyonicQuantumSpin2013}. In both cases we will assume open boundary conditions so that the lattice systems are described by a BCFT in the thermodynamic limit. The numerics will be used to verify our projectors \labelcref{eq:MFC-projector} and symmetry resolve the spectrum in the presence of boundaries.}

\JHc{As was reviewed in Section~\ref{sec:SREE-BCFT-review}, the entanglement spectrum of a CFT coincides with the spectrum of an associated BCFT \cite{cardyEntanglementHamiltoniansTwodimensional2016}. In this spirit, we will identify the Hamiltonian $H_{\mathrm{OBC}}$ of our open anyonic chains with the Hamiltonian of a BCFT and interpret it as describing the entanglement properties of another CFT. Moreover, we interpret the symmetry resolution of the energy spectrum of $H_{\mathrm{OBC}}$ as symmetry resolving the entanglement spectrum of that CFT.}

\TQc{To be specific, we can construct the `reduced density matrix' \labelcref{eq:rdm} associated with the BCFT  as \JHcc{$\rho_A \propto e^{-H_{\mathrm{OBC}}}$}.
This is an abuse of notation, as $H_{\mathrm{OBC}}$ is not a genuine entanglement Hamiltonian but rather a local BCFT Hamiltonian.
We use this notation to bypass the need for additional notation in our numerical analysis, where we work with local BCFT Hamiltonians, rather than the non-local entanglement Hamiltonians. 
Therefore, throughout this section on numerics, we will still refer to $\rho_A$ as the reduced density matrix, with the understanding that it is the local BCFT Hamiltonian which is exponentiated, rather than a genuine entanglement Hamiltonian.
}
 \subsubsection{Numerical check: golden chain}
 
 \begin{figure}[]
   \centering
 \includegraphics[]{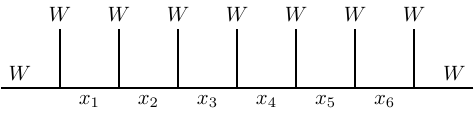}
   \caption{The golden chain with $ L = 6$ links and $ L_A = 7 $ external $ W$ anyons.
   The boundary anyons are fixed, $x_0 = x_{L+1} = W$.}
   \label{fig:golden-chain}
 \end{figure}
 
\JHc{Firstly, we study the golden chain.} This is an anyonic lattice model describing the tricritical Ising CFT \cite{feiguinInteractingAnyonsTopological2007} (see also \cite{Andrews:1984JSP....35..193A,Klumper:1992PhyA..183..304K}).
 We consider an open golden chain with $ L_A$ external anyons or $L = L_A - 1 $ links as illustrated in \cref{fig:golden-chain}.
 The local link variables take values in the Fibonacci category, $ x_i \in \{1, W\}$ for $i = 1, \dots, L$.
 The neighbouring links are subject to the constraints of the Fibonacci fusion rules and we fix the boundary anyons to be $ x_0 = x_{L+1} = W$.
 The Hamiltonian of the golden chain projects neighbouring anyons to fuse into the trivial channel.
 Using the local basis $ \{1, W \}$ for the link variables, the Hamiltonian is \cite{feiguinInteractingAnyonsTopological2007} 
 \begin{widetext}
 \begin{align}
   H_{\mathrm{Golden} } = - \sum_{i}^{} h_i \qquad\text{with} 
 \qquad h_i = n_{i-1} + n_{i+1} -n_{i-1}n_{i+1}\left(\varphi^{-3/2}+ \sigma_i^x + \varphi^{-3} n_i +1 + \varphi^{-2}\right)\;,
 \end{align}
 \end{widetext}
 where $ n_i$ counts the number of $ W$ particles at link $ i$ and $ \sigma^x_i$ is the Pauli $ x$ matrix.
 The spectrum of the open golden chain is described by the tricritical Ising model BCFT.
 For an even (odd) number of anyons, the spectrum is described by a BCFT with boundary conditions $ a = W$ and $ b = W$ ($ \eta W$) \cite{feiguinInteractingAnyonsTopological2007}.
 
 We determined the spectrum of $H_{\mathrm{Golden}}$ using exact diagonalization for chain lengths $L$ ranging from $3$ to $L=19$ and organized it in irreducible representations with respect to the Fibonacci symmetry.
 
 To achieve this, we constructed the Fibonacci projectors \eqref{eq:Tricrit-Project} on the lattice.
 This requires the lattice analogs of the Verlinde lines $ \mathcal{L}_1$ and $ \mathcal{L}_W$.
 \begin{figure*}[]
   \centering
   \includegraphics[]{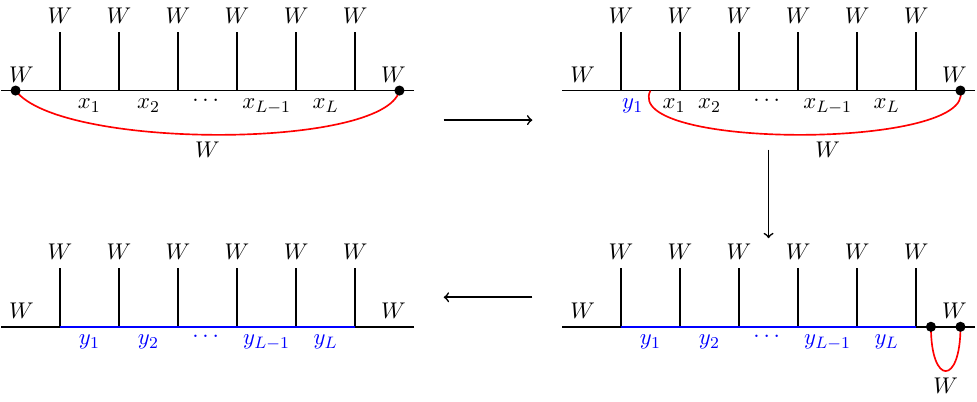}
   \caption{Schematics of the action of the topological symmetry on the anyonic chain.
   The filled circles denote the junction normalisation, see \cref{fig:junction-norm}.
 By performing a sequence of $ F$-moves, the defect can be transported across the anyonic chain and then shrunk at the boundary.}
   \label{fig:topo-sym}
 \end{figure*}
 The defects on the lattice, known as the topological symmetry \cite{feiguinInteractingAnyonsTopological2007}, can be described pictorially as illustrated in \cref{fig:topo-sym}.
 The general action of the topological symmetry $\mathcal{L}_i$ for boundary anyons $ x_0 = x_{L+1} = \mathcal{B} $ and external anyons $ j$ is
 \begin{align}
   \mathcal{L}_i& \ket{x_1,\dots,x_L} \nonumber \\
   &= \sum_{y_1,\dots,y_L}^{} \prod_{n = 0}^L F_{x_n y_{n+1}} \begin{bmatrix}
     y_n & j \\ i &  x_{n+1}
   \end{bmatrix}\ket{y_1, \dots, y_L} \;. \label{eq:golden-toposym}
 \end{align}
 Here, $ x_0 = x_{L+1} = y_0 = y_{L+1} = \mathcal{B} $ and $ i$ labels the defect $ \mathcal{\mathcal{L}}_i$.
 For the golden chain, the boundary anyons and external anyons are $\mathcal{B}   = j= W$.
 Note that \cref{eq:golden-toposym} differs from the topological symmetry in Ref.~\cite{choiRemarksBoundariesAnomalies2023} by a factor of $\sqrt{d_i}$ due to the presence of junction factors in \cref{fig:topo-sym}.
 
 Both boundaries in the golden chain are labelled by $W$ and therefore the non-trivial boundary fusion rule is given by \cref{eq:W-fusion} no matter the chain length.
 However, the above statement is not obvious from the BCFT perspective.
 This is because the boundary conditions in the BCFT describing the open golden chain differs depending on whether the number of external anyons is odd or even \cite{feiguinInteractingAnyonsTopological2007}.
 It is not obvious that the boundary fusion rules in \cref{eq:boundary-fusion-rules} should be identical when there is an even number of external anyons, $a = b = W$, and when there is an odd number of external anyons, $a = W$ and $b = \eta W$.
 For this to happen we must have
 \begin{align}
     \tilde{N}_{cd}^{[W] e} = \tilde{N}_{cd}^{[W, \eta W] e}\;,
 \end{align}
 for $c$, $d$, $e \in C=\{1, W\}$.
 This example is a lattice realisation showing that the boundary fusion rules of $C$ can be identical for different boundary conditions in the BCFT.
 
\begin{figure*}[]
   \centering
     \includegraphics[width=0.47\textwidth]{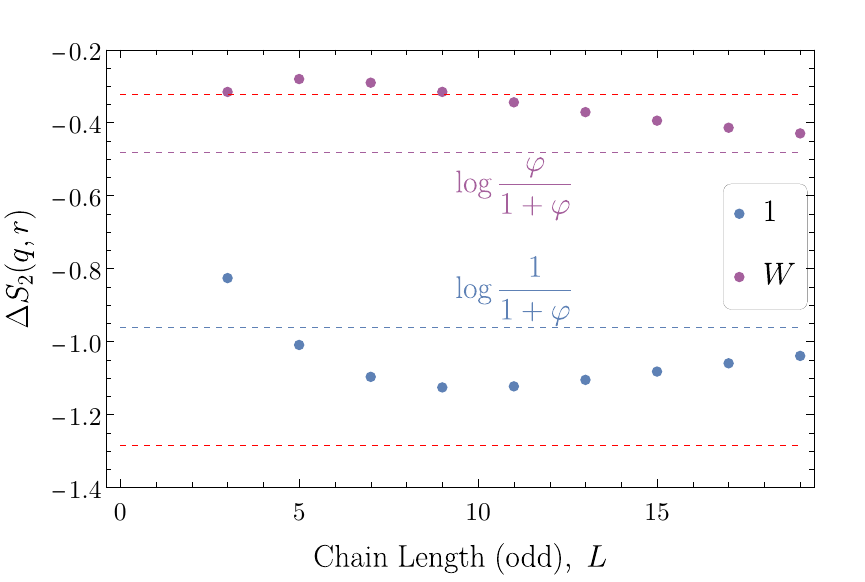}
     \hspace{0.5cm}
     \includegraphics[width=0.47\textwidth]{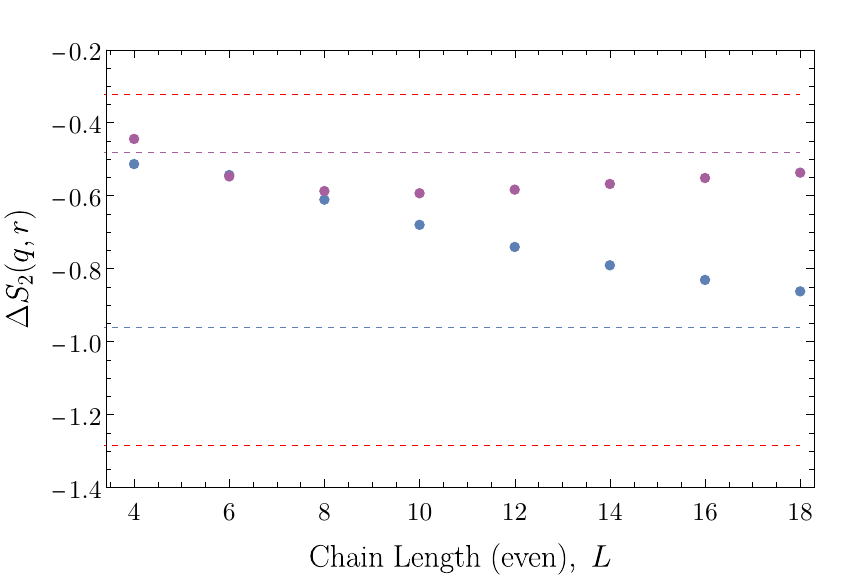}
     \caption{
     The legend labelling the irreps of the Fibonacci category is shown in the left panel.
     For both plots, $\Delta S_2(q,r)$, for $r \in \{1,W \}$, is shown for the number of links being (left) odd and (right) even. 
     The red dashes show the values predicted by \cite{saura-bastidaCategoricalsymmetryResolvedEntanglement2024}: $\log d_r^2/\abs{C}^2$, where $C = \{1,W\}$; explicitly, the dashes correspond to $ \log 1/(1+\varphi^2) \approx -1.286$ and $ \log \varphi^2/(1+\varphi^2) \approx -0.326$.}
   \label{fig:golden-chain-numerics}
 \end{figure*}

 The projectors onto the two irreps of the Fibonacci category are given by \cref{eq:Tricrit-Project}.
 Using the asymptotic equipartition of the SREE, \cref{eq:SREE-equipartition}, we expect
 \begin{align}
   \Delta S_2(q, 1) =  \log \frac{1}{1+\varphi}\;, \ \Delta S_2(q, W) =  \log \frac{\varphi}{1+\varphi}\;, \label{eq:expected-result}
 \end{align}
 where $ 1$ is the trivial irrep and $ W$ is the non-trivial irrep.
 \JHc{We compute \cref{eq:expected-result} by using $\rho_A \propto e^{-H_{\mathrm{Golden}}}$.}
 Our numeric results plotting $ \Delta S_2(q, r)$, $ r \in \{1, W \}$, are shown in \cref{fig:golden-chain-numerics}.
 We find that both the $ 1$ and $ W$ irreducible representation are approaching their predicted asymptotic value.
 This provides compelling evidence that the appropriate asymptotic values of the SREE for the Fibonacci category are given by \cref{eq:expected-result}.
 However, the anyonic chains we could simulate were not long enough to observe saturation of the asymptotic value.
 
 \subsection{$ \rm{Rep}{(S_3)}$} \label{sec:Rep(S3)}
 
 When constructing the BCFT projectors \eqref{eq:MFC-general?}, we did not need to assume that the symmetry category of interest, $C$, was modular.
 If one can find an appropriate embedding $C \subset \mathcal{M}$ of the symmetry category $C$ into a modular fusion category $\mathcal{M}$, then projectors for the fusion category $C$ can also be constructed.
 We will illustrate this procedure by constructing projectors for the Rep($S_3$) fusion category in the presence of boundaries.
 
 To construct these in the context of a CFT, we consider the diagonal, $ A$-type modular invariant for the $ SU(2)_4$ Wess-Zumino-Witten model.
 This theory contains fields with spin $ 0$, $ 1/2$, $ 1$, $ 3/2$ and $ 2$.
 As mentioned in \cite{choiRemarksBoundariesAnomalies2023}, the boundary state $ \ket{k/4 = 1}$ is weakly symmetric under a $ {C} =\rm{Rep}(S_3)$ category that is generated by three simple lines: $ \mathcal{L}_j$, for $ j =0$, $ 1$ and $ 2$.
  $ \mathcal{L}_0$ is the identity line and $ \mathcal{L}_1$ and $ \mathcal{L}_2$  satisfy the fusion rules
  \begin{align}
    2 \times 2 &= 0 \nonumber \\
    1 \times 1 &= 0 + 1 + 2 \nonumber \\
    1 \times 2 = 2\times 1 &= 1\;.
  \end{align}
  As expected, the above fusion rules show that the Cardy state $ \ket{ 1}$ is weakly symmetric with respect to $C$.
  The open string partition function with both boundaries fixed as $ \ket{1}$ is
  \begin{align}
    Z_{11}(q) &= \chi_0(q) + \chi_1(q) + \chi_2(q)\;.
  \end{align}
 The action of $ C$ on the boundary operators is described by the $ F$-symbol
  \begin{align}
    F_{11}\mqty[1  & d \\i & 1]\;,
  \end{align}
  for $ d,i \in {C}$.
  The relevant, non-trivial $ F$-symbols are
  \begin{align}
    F_{11}\mqty[1 & 1 \\2 & 1] = -1\;, \quad F_{11}\mqty[1 & 2 \\1 & 1] = -1\;, \nonumber \\ 
    F_{11}\mqty[1 & 1 \\1 & 1] = 0\;. \label{eq:Rep(S3)-F}
  \end{align}
  Using \cref{eq:BCFT-projector}, we find the three projectors 
  \begin{align}
    \Pi_0^{[1]} &= P_0^{[1]} = \frac{1}{4}\qty( \mathcal{L}_0 + 2 \mathcal{L}_1 + \mathcal{L}_2) \nonumber \\
    \Pi_1^{[1]}  &= P_1^{[1]}= \frac{1}{4}\qty(2 \mathcal{L}_0  - 2 \mathcal{L}_2) \label{eq:Rep(S3)-projectors} \\
    \Pi_2^{[1]}  &= P_2^{[1]}= \frac{1}{4}\qty(\mathcal{L}_0 - 2 \mathcal{L}_1 + \mathcal{L}_2)\;. \nonumber
  \end{align}
  These projectors imply that for a Rep($ S_3$) symmetry the equipartition of the SREE is broken by, \cref{eq:SREE-equipartition},
  \begin{align}
    \Delta S_n(q, 0) = \Delta S_n(q,2) = \log \frac{1}{4}\;, \qquad \Delta S_n(q,1) = \log \frac{1}{2}\;.
  \end{align}
  Using the projectors \eqref{eq:Rep(S3)-projectors} along with the non-trivial $ F$-symbols \eqref{eq:Rep(S3)-F}, a quick calculation shows
  \begin{align}
    \tr_{22} \Pi_0^{[1]} \rho_{A,11} &= \chi_0(q)\;, \nonumber \\ \tr_{22}\Pi_1^{[1]} \rho_{A,11} &= \chi_1(q)\;, \nonumber \\ \tr_{22} \Pi_2^{[1]}\rho_{A,11} &= \chi_2(q)\;.
  \end{align}
   The boundary fusion rules of Rep($ S_3$) are identical to the bulk fusion rules, with the peculiar exception
  \begin{align}
    \mathcal{L}_1^{[1]} \times \mathcal{L}_1^{[1]} = \frac{1}{2}\mathcal{L}_0^{[1]} + \frac{1}{2}\mathcal{L}_2^{[1]}\;. \label{eq:Rep(S3)-exception}
  \end{align}
  Note that no non-invertible symmetry appears on the right hand side of \cref{eq:Rep(S3)-exception}; this is to be contrasted with the bulk fusion rule $ 1 \times 1 = 0 + 1 + 2$. 
  Using these fusion rules, one can show that the projectors \eqref{eq:Rep(S3)-projectors} are orthogonal projectors.
  
 \subsubsection{Numerical check: $ SU(2)_4$ spin-1/2 anyon chain} \label{sec:su(2)4-anyon} 
\begin{figure}
   \centering
   \includegraphics[]{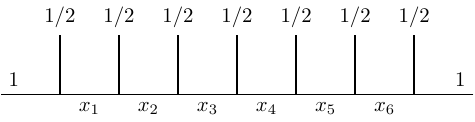}
   \caption{The $ SU(2)_4$ anyon chain with external spin-1/2 anyons and fixed spin-1 anyons at the boundary.}
   \label{fig:SU(2)4-chain}
 \end{figure}

 To study the Rep($ S_3$) projectors \eqref{eq:Rep(S3)-projectors}, we numerically simulate the $ SU(2)_4$ spin-1/2 antiferromagnetic anyonic chain.
\JHc{ Our methodology is identical to our study of the golden chain, and is described in \cref{sec:num-general}.}
 This chain describes the Tetracritical Ising model \cite{feiguinInteractingAnyonsTopological2007,gilsAnyonicQuantumSpin2013}, which has a Rep($ S_3$) symmetry.
 We study an open chain with spin-1/2 external anyons and spin-1 anyons as boundaries, see \cref{fig:SU(2)4-chain} for an illustration.
 The Hamiltonian favours neighbouring spin-1/2 anyons to fuse into the trivial channel.
 Due to the fusion rules, the local basis alternates between integer labels and half-integer labels.
 Therefore, the number of links $L$ on the anyon chain must be odd because we have fixed spin-1 anyons on the boundary.
 In the local basis $ x_{i} \in \{1/2, 3/2 \}$, $ x_{i+1} \in \{0, 1, 2 \}$ where $ i$ $ (i+1)$ is odd (even), the Hamiltonian is 
 \begin{widetext}
 \begin{align}
   H_{SU(2)_4} &= - \sum_{i}^{} h_i \text{ with }   \label{eq:SU(2)4-Ham}\\
   h_i &= n_{i-1}^0 n_{i+1}^0 + n_{i-1}^1 n^1_{i+1} \begin{bmatrix}
     1 & -1 \\ -1 & 1
   \end{bmatrix}_i
   + \frac{1}{3} n_{i}^{1/2} n_{i}^{1/2} \begin{bmatrix}
     1 & -\sqrt{2} \\ -\sqrt{2}  & 2
   \end{bmatrix}_{i+2}
   +\frac{1}{3} n_{i}^{3/2} n_{i}^{3/2} \begin{bmatrix}
     2 & -\sqrt{2}  \\ - \sqrt{2}  & 1
   \end{bmatrix}_{i+2} \;.  \nonumber
 \end{align}
 \end{widetext}
 
\begin{figure*}[]
   \centering
     \includegraphics[width=0.47\textwidth]{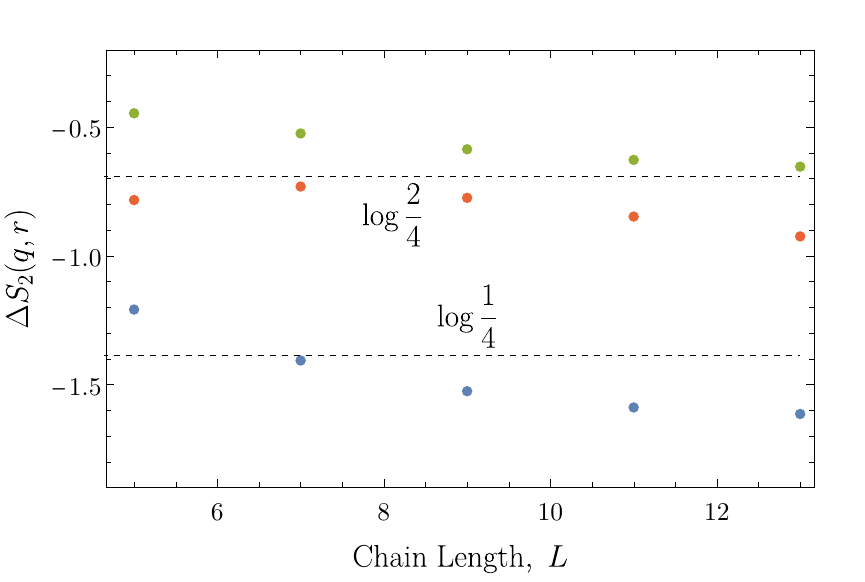}
     \hspace{0.5cm}
     \includegraphics[width=0.47\textwidth]{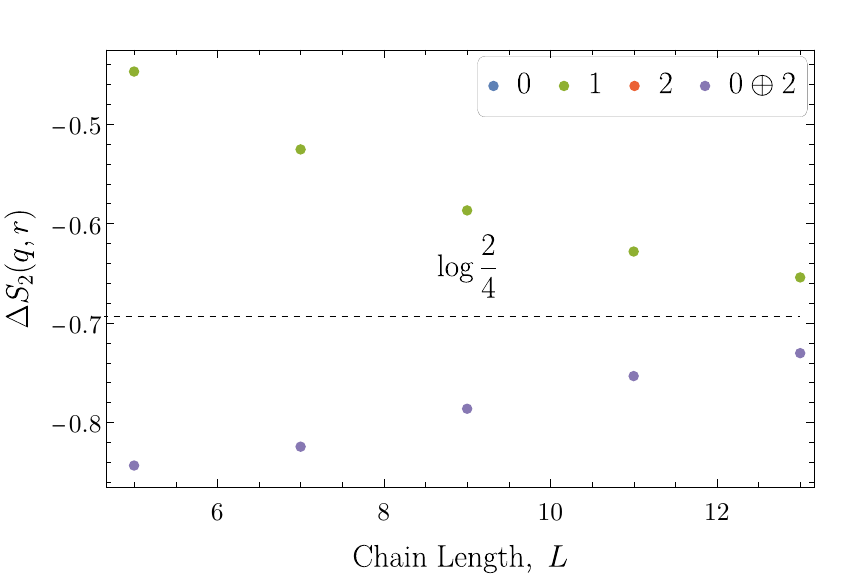}
     \caption{The legend for both plots, denoting the irreps of Rep($S_3$), is shown in the top right corner of the right panel. (Left) $ \Delta S_2(q,r)$ shown for $ r \in\{0,1,2 \}$. 
   (Right) $ \Delta S_2(q, 0 \oplus 2)$ and $ \Delta S_2(q,1)$ are shown.}
   \label{fig:Rep(S3)-numerics}
 \end{figure*}

 A lattice realisation of the topological symmetry is analogous to the case of the golden chain in \cref{eq:golden-toposym}.
 The lattice realisation of the defects is given by \cref{eq:golden-toposym} with the boundaries $ x_0 = x_{L+1} = y_0 = y_{L+1} = \mathcal{B} = 1$, and external anyons $ j = 1/2$.
 Via arguments similar to the case of the golden chain, it is clear that the fusion rules of the $ \mathcal{L}_i$ are identical to those in (and above) \cref{eq:Rep(S3)-exception}.
 Therefore, we can expect the following equipartition of the SREE:
 \begin{align}
   \Delta S_n(q,0) = \Delta S_n(q,2) = \log \frac{1}{4}\;, \qquad \Delta S_n (q, 1) = \log \frac{1}{2}\;.
 \end{align}
 Similar to the golden chain, we define the reduced density matrix as $ \rho_A\coloneq e^{-H_{SU(2)_4}}$.
 We then compute $ \Delta S_2(q, r)$ for various chain lengths via exact diagonalisation and the Rep($S_3$) projectors \labelcref{eq:Rep(S3)-projectors}.
 \cref{fig:Rep(S3)-numerics}(left) shows $ \Delta S_2(q,r)$ for $ r \in \{0,1,2\}$ for various chain lengths. 
 While the spin-$2$  representation appears to be converging to $ \log 1/2$, the spin-$0$ and spin-$1$ representations are far from the expected value of $ \log 1/4$.
 We suspect that the difference in the rate of convergence between the spin-$2$ representation and the spin-$0$ and spin-$1$ representation is due to the number of states in each representation.
 In the anyonic chain Hilbert space, approximately half of the states transform in the $ 1$ representation, while only one quarter of the states transform in the $ 0$ or $ 2$ representation. This follows from investigating the rank of the respective projection operators but it also matches the corresponding ratios of quantum dimensions, $d_1/d_i = 2$ for $i = 0$ and $2$.
 Thus one would expect $ \Delta S_n(q,2)$ to converge faster on the lattice as it is, in a sense, closer to the thermodynamic limit.
 
 With this in mind, observe that the number of states in the combined $0 \oplus 2$ representation is equal to the number of states in the $1$ representation.
 We then expect that $ \Delta S_2(q, 0 \oplus 2)$ will converge at approximately the same rate as $ \Delta S_2(q, 1)$ and the expected limit will be
 $ \Delta S_2(q, 0 \oplus 2 ) = \log 1/2$.
 Indeed, as shown in \cref{fig:Rep(S3)-numerics}(right), $\Delta S_2(q, 0\oplus 2)$ and  $\Delta S_2(q, 1)$ approach the expected value of $ \log 1/2$ at similar rates as the chain length is increased.
 However, similar to the case of the golden chain, the length of the chain $L$ is not long enough for this asymptotic value to be saturated.
 
 \subsection{Additional observations} 
 
 After studying numerous examples in the previous sections, we have arrived at the following observation concerning the asymptotic limit of the SREE, \cref{eq:asymptotic-1}.
 Firstly, when we used the BCFT projectors $P_r^{[a,b]}$ \labelcref{eq:BCFT-projector}, recall that the asymptotic symmetry resolved partition function \labelcref{eq:asymptotic-1} contained the prefactor $d_r/d_a d_b$, where $a$ and $b$ are the (simple) boundaries at the entangling surface and $r \in a \times b$ labels a boundary conformal tower.
 At first glance, this factor appears to have no relation to the set of Verlinde lines that the boundaries $a$ and $b$ are weakly symmetric under, $\mathrm{WS}^{[a,b]}$.
 However, we believe that 
 \begin{align}
     \frac{d_r}{d_a d_b} = \frac{d_{c}}{\sqrt{\mathrm{WS}^{[a,b ]}}}\;, \label{eq:asymptotic-conjecture}
 \end{align}
 for some $c \in \mathrm{WS}^{[a,b]}$ and $\sqrt{\mathrm{WS}^{[a,b]}} = \sum_{c \in \mathrm{WS}^{[a,b]}} d_c$.
 In other words, the projectors onto the conformal families, $P_r^{[a,b]}$, project onto the irreps of $\mathrm{WS}^{[a,b]}$.
 This provides an interpretation of the asymptotic limit \labelcref{eq:asymptotic-1} in terms of the Verlinde lines present in the open string Hilbert space.
 For the case where $a = b$, \cref{eq:asymptotic-conjecture} is easily seen to be true, with $c = r$ and $d_a^2 = \sqrt{\mathrm{WS}^{[a,b]}}$.
 More generally, for $a \neq b$ it is not obvious that \cref{eq:asymptotic-conjecture} is true, although we have checked it for some specific minimal models, such as the tricritical and tetracritical Ising model.
 
 Let us briefly discuss the reasoning behind \cref{eq:asymptotic-conjecture} in the general case, $a \neq b$.
 Firstly, by using the Verlinde formula, one can show that
 \begin{align}
     \sum_{d} N_{ad}^a N_{bd}^b = \sum_{r} N_{ab}^r\;, \label{eq:N-equality}
 \end{align}
 which implies that the number of boundary conformal towers in the partition function $Z_{ab}(q)$ is equal to the number of simple objects in $\mathrm{WS}^{[a,b]}$.
 Then, if the irreducible representations of $\mathrm{WS}^{[a,b]}$ are in one-to-one correspondence with the boundary conformal towers i.e. for every $r \in a \times b$ there is a corresponding irrep $c \in \mathrm{WS}^{[a,b]}$,
 the projector onto the boundary conformal tower $r$, $P_{r}^{[a,b]}$, then projects onto the irreducible representation $c \in \mathrm{WS}^{[a,b]}$.
 Appealing to the specific case of both boundaries being equal \labelcref{eq:MFC-SREE-equipartition}, we then suggest the relation in \cref{eq:asymptotic-conjecture}.
 
 \subsection{Boundary conditions and projectors} \label{sec:Boundary-Independence}
 
 It is worth noting that the projectors we found for the Fibonacci symmetry in the three-state Potts model, \cref{eq:Potts-Fib-Projectors}, are identical to the Fibonacci projectors for the tricritical Ising model, \cref{eq:Tricrit-Project}.
 Naturally, this leads one to ask: do the projectors defined in \cref{eq:BCFT-projector} actually depend on the choice of boundary conditions?
 For the case of the Fibonacci category, one can show the BCFT projectors do not depend on the choice of boundary condition when they are equal on both sides.
 Suppose we choose both boundary conditions to be labelled by $ a$;
 the boundary fusion algebra is 
 \begin{align}
   \mathcal{L}_c^{[a]} \times \mathcal{L}_d^{[a]} = \sum_{e \in c \times d}^{} \tilde{N}_{cd}^{[a] e} \mathcal{L}_e^{[a]}\;,
 \end{align}
 where $ N_{cd}^{[a] e}$ need not be integers, see \cref{eq:boundary-fusion-rules}.
 The boundary fusion algebra identities \labelcref{eq:boundary-fusion-identities} completely determine the boundary fusion rules for the Fibonacci category:
 \begin{align}
   \mathcal{L}_W^{[a]} \times \mathcal{L}_W^{[a]} = \frac{1}{\varphi} \mathcal{L}_1^{[a]} + \frac{\varphi - 1}{\varphi} \mathcal{L}_W^{[a]}\;.
 \end{align}
 The orthogonal projectors that are compatible with these fusion rules are
 \begin{align}
   \Pi_1^{[a]} &= \frac{1}{1+\varphi}(\mathcal{L}_1 + \varphi \mathcal{L}_W) \;, \nonumber \\
   \Pi_{W}^{[a]} &= \frac{\varphi}{1+\varphi}(\mathcal{L}_1 - \mathcal{L}_W)\;.
 \end{align}
 
 Finally, from the perspective of critical anyonic chains, it is somewhat natural to conclude that the boundary fusion rules are independent of the specific choice of boundaries.
 We briefly discuss this for the case of Rep($ S_3$) symmetry.
 
 There are numerous anyonic chains with Rep($ S_3$) symmetry with open boundary conditions.  For instance, the (anti) ferromagnetic $ SU(2)_4$ spin-1/2 chain describes the (Tetracritical Ising) $ Z_4$ parafermion CFT \cite{gilsAnyonicQuantumSpin2013}.
 Similarly, the $ SU(2)_4$ spin-1 anyon chains can describe $ c = 1$ compact boson CFTs at various compactification radii \cite{gilsAnyonicQuantumSpin2013} (including non-rational points).
 In the above cases, fixing a spin-1 anyon on the boundaries of the open chain will result in the Rep($ S_3$) symmetry being present.
 The boundary fusion rules for Rep($ S_3$) will be those stated in \cref{eq:Rep(S3)-exception}.
 Therefore, from the lattice perspective, it is obvious that with spin-1 anyons at the boundaries the Rep($ S_3$) projectors \eqref{eq:Rep(S3)-projectors} will project onto the irreps of Rep($ S_3$).
 
 In contrast, from the continuum BCFT perspective, it is non-trivial to determine whether these projectors are also the appropriate projectors for Rep($S_3$).
 Firstly, this is because the BCFTs associated with the anyonic chains will generally admit non-Cardy boundary states.
 Secondly, as shown for the golden chain, even if both boundary anyons are identical, the corresponding boundary states in the BCFT can be different.
 For these reasons, it is not obvious that the boundary fusion rules in the presence of these boundaries are given by \cref{eq:Rep(S3)-exception} in the BCFT describing the continuum limit.
 However, viewing these BCFTs as the continuum limit of the associated anyonic chains, the boundary fusion rules of the Rep($ S_3$) symmetry in the continuum must also satisfy \cref{eq:Rep(S3)-exception}.
 This would suggest that as long as the boundary conditions in the CFT are weakly symmetric under Rep($ S_3$), then fusion of Rep($ S_3$) in the presence of these boundaries is identical to \cref{eq:Rep(S3)-exception}.
 Therefore, we tentatively suggest the following picture. Suppose that the simple boundary conditions $ a$ and $ b$ are weakly symmetric under a symmetry category $ C$ that acts faithfully (see \cite{linAsymptoticDensityStates2023}) on the Hilbert space of interest.
 Then the fusion of objects in $ C$ in the presence of the boundaries $ a$ and $ b$ are independent of the specific choice of $ a$ and $ b$.
 
 \section{Conclusions} 
 In this work we have analysed the symmetry-resolved entanglement entropy for non-invertible symmetries.
 Specifically, we focused on non-invertible symmetries generated by Verlinde lines in $1{+}1$d conformal field theories with diagonal partition function.
 We constructed projectors onto irreducible  sectors for these non-invertible symmetries in the presence of boundaries and utilised them to study the equipartition of the symmetry-resolved entanglement entropy. A key element was the study of the defect fusion algebra in the presence of boundaries which permits non-integer coefficients.
 Explicit examples of orthogonal projectors for the cases of the Fibonacci category and Rep($ S_3$) were provided.
 Importantly, the Fibonacci projectors and the associated expressions for the symmetry-resolved entanglement entropy differ from those in the recent literature \cite{saura-bastidaCategoricalsymmetryResolvedEntanglement2024}.
 
 An obvious direction for future work is extending the analysis to general fusion categories whose fusion rules may be non-commutative. This is likely best achieved by utilising the SymTFT formalism \cite{gaiottoOrbifoldGroupoids2021, apruzziSymmetryTFTsString2023, freedTopologicalSymmetryQuantum2024}.
 Indeed, the SymTFT proved to be a powerful tool when studying the symmetry resolution of the torus partition function \cite{linAsymptoticDensityStates2023}.
 This is because the SymTFT can be used to construct the generalised charges associated with a symmetry \cite{bhardwajGeneralizedChargesPart2024, bhardwajGeneralizedChargesPart2023}.
 Work in this direction requires further understanding of the representation theory of non-invertible symmetries on manifolds with boundaries, which has been recently studied in the context of the representation theory of solitons \cite{Copetti:2024dcz, Copetti:2024onh, cordovaRepresentationTheorySolitons2024}.
 
 Throughout this work we have emphasised that the fusion algebra of non-invertible symmetries in the presence of boundaries is different than the bulk fusion algebra. This leads to a natural question: does the fusion algebra explicitly depend on the choice of boundary conditions?
 We commented on this briefly in \cref{sec:Boundary-Independence} and tentatively suggest that the boundary fusion relations are independent of the choice of boundary conditions, as long as the boundaries are weakly symmetric under the fusion category of interest.
 While one may have physical reasons to believe this picture to be true, a mathematical proof is far from obvious as the boundary fusion algebra makes explicit reference to the choice of boundary conditions.
 
 Additionally, it would be beneficial to further analyse the interplay between entanglement measures and non-invertible symmetries.
 Entanglement measures require some choice of bipartition and consequently a choice of boundary conditions at the entangling surface.
 Therefore, we believe our results will prove useful in analysing the relation between other entanglement measures and non-invertible symmetries, such as the entanglement asymmetry \cite{aresEntanglementAsymmetryProbe2023, capizziEntanglementAsymmetryOrdered2023, capizziUniversalFormulaEntanglement2024, chenEnyiEntanglementAsymmetry2024, fossatiEntanglementAsymmetryCFT2024}.
 On a similar note, to study these entanglement measures numerically requires realising the non-invertible symmetries on a lattice with open boundary conditions.
 While recent works have constructed Rep($ S_3$) symmetries for periodic chains \cite{chatterjeeQuantumPhasesTransitions2024, bhardwajIllustratingCategoricalLandau2024}, they have, to our knowledge, not yet been constructed on spin chains with open boundary conditions. 
 We leave the discussion of spin chain realisations of non-invertible symmetries in the presence of boundaries to future work.
 
 \emph{Note added.} While in the process of completing this work we became aware of related work on symmetry-resolved entanglement measures for non-invertible symmetries by Yichul Choi, Brandon Rayhaun, and Yunqin Zheng \cite{Zheng2} as well as Arpit Das, Javier Molina-Vilaplana and Pablo Saura-Bastida \cite{Das}. We also expect parts of other upcoming work on the SymTFT picture for physical theories with boundaries by Yichul Choi, Brandon Rayhaun, and Yunqin Zheng \cite{Zheng1}, Lakshya Bhardwaj, Christian Copetti, Daniel Pajer, and Sakura Sch\"afer-Nameki \cite{Bhardwaj} as well as I\~naki Garc\'ia Etxebarria, Jes\'us Huertas, and Angel Uranga \cite{GarciaEtxebarria} to be relevant in our context.
 
 \subsection*{Acknowledgement}
   The research of JH was supported by an Australian Government Research Training Program (RTP) Scholarship.
   The authors thank Tyler Franke for helpful discussions and Hong-Hao Tu for bringing the concept of a ladder algebra to their attention. 
   We would like to thank everyone listed under ``Note added'' for willing to coordinate our respective submissions.


\appendix
\section{Projector calculations}  

\subsection{Tetrahedron symbols}\label{sec:tet-symbols}

The tetrahedron symbols are obtained by considering a defect network in the shape of a tetrahedron.
Following the conventions in Ref.~\cite{kojitaTopologicalDefectsOpen2018} and using \cref{eq:isotopy-invariant}, they are related to the $F$-symbols by
\begin{align}
\begin{tikzpicture}[baseline={(0, 0cm-\MathAxis pt)}] 
  \draw[black,thick] (0,0) circle (1);
  \draw[black,thick] (0,0) -- node[midway, anchor =east] {$\tiny{i}$} (0,1);
  \draw[black,thick] (0,0) -- node[midway, anchor =south west, xshift=-0.1cm] {$\tiny{j}$}(0.866,-0.5);
  \draw[black,thick] (0,0) -- node[midway, anchor =north west , xshift=-0.1cm, yshift=0.1cm] {$\tiny{k}$}(-0.866,-0.5);
 \draw (45: 1) node[anchor=south west]{$c$};
 \draw (-90: 1) node[anchor=north]{$a$};
 \draw (135: 1) node[anchor=south east]{$b $};
\end{tikzpicture}
\quad &=
  \begin{bmatrix}
    i & j & k \\ a & b & c
  \end{bmatrix}^{\mathrm{TET} } \nonumber \\ 
  &= \sqrt{d_a d_b d_i d_j} F_{ck} \begin{bmatrix}
    j & i \\a & b
  \end{bmatrix}\;. \label{eq:tet-symbols}
\end{align}
There are numerous ways to shrink this tetrahedron to a point, which generates numerous identities between the $ F$-symbols.
\cref{eq:isotopy-invariant} becomes more useful when used in conjunction with the tetrahedron symbols,
Due to the numerous way to shrink a tetrahedron to a point, \cref{eq:tet-symbols} has many symmetries.
For instance, \cref{eq:tet-symbols} is invariant under the cyclic permutation of its columns
\begin{align}
   \begin{bmatrix}
    i & j & k \\ a & b & c
  \end{bmatrix}^{\mathrm{TET} } 
  = \begin{bmatrix}
    k & i & j \\ c & a & b
  \end{bmatrix}^{\mathrm{TET} }
  = \begin{bmatrix}
    j & k & i \\ b & c & a
  \end{bmatrix}^{\mathrm{TET} }\;,
\end{align}
or switching the upper and lower labels of two different columns simultaneously.

\subsection{Pentagon identity} \label{sec:pentagon-id-proj}

The pentagon identity \eqref{eq:pentagon-identity} reads 
\begin{align}
  \sum_{s }^{} F_{ps} \mqty[b & c \\ a & q] F_{qt}\mqty[s & d \\ a & e] F_{sr}\mqty[c & d \\ b & t] &= F_{qr}\mqty[c & d \\p & e] F_{pt}\mqty[b & r \\ a & e]\;.
\end{align}
If $ p = 1$ then we require $ a = b$, $ c = q$ and $ e = r$.
Then,  by the assumption of parity invariance of defects, $ F_{pq}[j k; il] = F_{pq}[i l; j k]$,\footnote{This is obtained by reflecting \cref{fig:F-symbols} about the horizontal.} we find
\begin{align}
  \sum_{s \in C}^{} F_{1s} \mqty[b & c \\ b & c] F_{ct}\mqty[b & e \\ s & d] F_{sr}\mqty[c & d \\ b & t] &=  F_{1t}\mqty[b & r \\ b & r] \delta_{er}\;. \label{eq:projector-proof-start}
\end{align}
Now we set $ b =c=X$ and $ t = d = Y$
\begin{widetext}
\begin{align}
    \sum_{s\in C}^{} F_{1s} \mqty[X & X \\ X & X] F_{XY}\mqty[X & e \\ s & Y] F_{sr}\mqty[X & Y \\ X & Y] &=  F_{1Y}\mqty[X & r \\ X & r] \delta_{er} \\
      \frac{1}{F_{1Y}\mqty[X & r \\ X &r]} \sum_{s}^{} F_{1s} \mqty[X & X \\ X & X]  F_{sr}\mqty[X & Y \\ X & Y] F_{XY}\mqty[X & e \\ s & Y] &= \delta_{re}\;. \label{eq:penatgon-proj-fin}
\end{align}
\end{widetext}
We identify the final term on the left side as the action of $ \mathcal{L}_s$ on the irreducible representation $ e$ with boundaries $ X$ and $ Y$;
this is only non-zero if the boundaries $ X$ and $ Y$ are weakly symmetric with respect to $ \mathcal{L}_s$.
Thus we arrive at 
\begin{align}
  P_r^{[a,b]} &= \frac{1}{F_{1b}\mqty[a & r \\a &r ]} \sum_{s\in \rm{WS}_{[a,b]}}^{} F_{1s}\mqty[a & a \\\ a & a] F_{sr}\mqty[a & b \\ a & b] \mathcal{L}_s\;,
\end{align}
as required.

\subsection{Orthogonal projectors}  \label{sec:appendix-projector-calc}

Let $C \subseteq \mathcal{M}$ where $\mathcal{M}$ is a modular fusion category.
Consider $ a \in \mathcal{M} $ such that $ a\times a = \sum_{c\in C}^{} C$.
Then using \cref{eq:BCFT-projector} we obtain the projector onto each representation $ r \in C$:
\begin{align}
  P_r^{[a]} &= \frac{1}{F_{1a} \begin{bmatrix}
      a & r \\ a & r
  \end{bmatrix}}
  \sum_{s \in C }^{} F_{1s} \begin{bmatrix}
    a & a \\ a & a
  \end{bmatrix}
  F_{sr} \begin{bmatrix}
     a& a \\ a & a
  \end{bmatrix}
  \mathcal{L}_s \nonumber \\
  &= \frac{\sqrt{d_r} }{d_a}  \sum_{s\in C}^{} \sqrt{d_s}  F_{sr} \begin{bmatrix}
    a & a \\ a & a
  \end{bmatrix}
  \mathcal{L}_s\;.
\end{align}
Next, using the $ F$-symbol identity (which is easily seen from the tetrahedron symbols)
\begin{align}
d_a   F_{sr} \begin{bmatrix}
    a & a \\ a & a
  \end{bmatrix} &=  \sqrt{d_s d_r}  F_{aa} \begin{bmatrix}
  a & r \\ s & a
  \end{bmatrix}
\end{align}
we then have
\begin{align}
  P_r^{[a]} &= \frac{d_r}{\sqrt{C} } \sum_{s \in C}^{} d_s F_{aa} \begin{bmatrix}
    a & r \\s & a
  \end{bmatrix}
  \mathcal{L}_s   
\end{align}
where we have used $ d_a^2 = \sqrt{\mathcal{M} } $. 
This recovers \cref{eq:MFC-general?}.
Next we show that these are orthogonal projectors.
We then compute
\begin{widetext}
\begin{align}
   P_r^{[a]} P_h^{[a]}  &= \sum_{s,d,\lambda}^{} N_{sd}^\lambda \tilde{N}_{sd}^{[a] \lambda}  \frac{F_{1s} \begin{bmatrix}
    a & a \\ a & a
  \end{bmatrix}
  F_{sr} \begin{bmatrix}
     a& a \\ a & a
  \end{bmatrix}
F_{1d} \begin{bmatrix}
    a & a \\ a & a
  \end{bmatrix}
  F_{dh} \begin{bmatrix}
     a& a \\ a & a
  \end{bmatrix}
}{ F_{1a} \begin{bmatrix}
  a & r \\a & r 
\end{bmatrix}
F_{1a} \begin{bmatrix}
  a & h \\ a & h
\end{bmatrix} }
\mathcal{L}_\lambda \nonumber \\
    &= \frac{\sqrt{d_r d_h} }{d_a^2} \sum_{s,d,\lambda}^{} \sqrt{d_s d_d}  N_{sd}^\lambda  {
  F_{sr} \begin{bmatrix}
     a& a \\ a & a
  \end{bmatrix}
  F_{dh} \begin{bmatrix}
     a& a \\ a & a
  \end{bmatrix}
}
 F_{\lambda a} \begin{bmatrix}
   s &d \\ a & a
 \end{bmatrix}
 F_{\lambda a} \begin{bmatrix}
   s & d \\ a & a
 \end{bmatrix}
 \mathcal{L}_\lambda \nonumber
\end{align}
\end{widetext}
From the above, by continually utilising the tetrahedron symbols and pentagon identity, one can show $P_r^{[a]} P_h^{[a]} = \delta_{rh} P_h^{[a]}$. 


\input{bib_bbl}
\end{document}

%% file: symbols.tex
\DeclarePairedDelimiter\ishiket{\lvert}{\rrangle}
\DeclarePairedDelimiterX\ishibraket[2]{\llangle}{\rrangle}{#1\,\delimsize\vert\,\mathopen{}#2}



%% file: bib_bbl.tex
\providecommand{\noopsort}[1]{}\providecommand{\singleletter}[1]{#1}%